\documentclass[reprint,
superscriptaddress,
 amsmath,amssymb,
 aps,
pra,
]{revtex4-2}
\usepackage{graphicx}% Include figure files
\usepackage{dcolumn}% Align table columns on decimal point
\usepackage{bm}% bold math
\usepackage{tensor}
\usepackage[dvipsnames]{xcolor}
\usepackage{soul}
\usepackage{svg}
\usepackage{subcaption}
\usepackage{tabularray}
\usepackage{physics}
\usepackage[utf8]{inputenc} % usually not needed (loaded by default)
\usepackage[T1]{fontenc}
\usepackage{dsfont}
\usepackage[symbol]{footmisc}
\usepackage[%  
    colorlinks=true,
    pdfborder={0 0 0},
    linkcolor=blue
]{hyperref}
\usepackage{url}
\usepackage{soul}
\begin{document}
\setstcolor{red}
\newcommand{\HY}[1]{{\color{blue}{[HY: #1]}}}
\newcommand{\XW}[1]{{\color{orange}{[XW: #1]}}}
\newcommand{\ANC}[1]{{\color{red}{[ANC: #1]}}}
\preprint{APS/123-QED}
%\title{Realizing tunable all-to-all connectivity using an on-chip quantum router}
%\title{Tunable coupler with all-to-all connectivity for modular superconducting quantum processors}
% \title{Realizing tunable all-to-all connectivity for modular superconducting quantum processors}
\title{Modular quantum processor with an all-to-all reconfigurable router}
\author{Xuntao Wu}
\thanks{These two authors contributed equally.}
\affiliation{Pritzker School of Molecular Engineering, University of Chicago, Chicago, IL 60637, USA}

\author{Haoxiong Yan}
\thanks{These two authors contributed equally.}
\affiliation{Pritzker School of Molecular Engineering, University of Chicago, Chicago, IL 60637, USA}

\author{Gustav Andersson}
\affiliation{Pritzker School of Molecular Engineering, University of Chicago, Chicago, IL 60637, USA}

\author{Alexander Anferov}
\affiliation{Pritzker School of Molecular Engineering, University of Chicago, Chicago, IL 60637, USA}
%\affiliation{Department of Physics, University of Chicago, Chicago, IL 60637, USA}

\author{Ming-Han Chou}
\altaffiliation[Present address: ]{AWS Center for Quantum Computing, Pasadena, CA 91125, USA}
\affiliation{Pritzker School of Molecular Engineering, University of Chicago, Chicago, IL 60637, USA}
\affiliation{Department of Physics, University of Chicago, Chicago, IL 60637, USA}

\author{Christopher R. Conner}
\affiliation{Pritzker School of Molecular Engineering, University of Chicago, Chicago, IL 60637, USA}

\author{Joel Grebel}
\altaffiliation[Present address: ]{Ecole Normale Superieure de Lyon, CNRS, Laboratoire de Physique, F-69342 Lyon, France}
\affiliation{Pritzker School of Molecular Engineering, University of Chicago, Chicago, IL 60637, USA}

\author{Yash J. Joshi}
\affiliation{Pritzker School of Molecular Engineering, University of Chicago, Chicago, IL 60637, USA}

\author{Shiheng Li}
\affiliation{Pritzker School of Molecular Engineering, University of Chicago, Chicago, IL 60637, USA}
\affiliation{Department of Physics, University of Chicago, Chicago, IL 60637, USA}

\author{Jacob M. Miller}
\affiliation{Pritzker School of Molecular Engineering, University of Chicago, Chicago, IL 60637, USA}
\affiliation{Department of Physics, University of Chicago, Chicago, IL 60637, USA}

\author{Rhys G. Povey}
\affiliation{Pritzker School of Molecular Engineering, University of Chicago, Chicago, IL 60637, USA}
\affiliation{Department of Physics, University of Chicago, Chicago, IL 60637, USA}

\author{Hong Qiao}
\affiliation{Pritzker School of Molecular Engineering, University of Chicago, Chicago, IL 60637, USA}

\author{Andrew N. Cleland}
\email{anc@uchicago.edu}
\affiliation{Pritzker School of Molecular Engineering, University of Chicago, Chicago, IL 60637, USA}
\affiliation{Center for Molecular Engineering and Material Science Division, Argonne National Laboratory, Lemont, IL 60439, USA}

\date{\today}

\begin{abstract}
Superconducting qubits provide a promising approach to large-scale fault-tolerant quantum computing. However, qubit connectivity on a planar surface is typically restricted to only a few neighboring qubits. Achieving longer-range and more flexible connectivity, which is particularly appealing in light of recent developments in error-correcting codes, however usually involves complex multi-layer packaging and external cabling, which is resource-intensive and can impose fidelity limitations. Here, we propose and realize a high-speed on-chip quantum processor that supports reconfigurable all-to-all coupling with a large on-off ratio. We implement the design in a four-node quantum processor, built with a modular design comprising a wiring substrate coupled to two separate qubit-bearing substrates, each including two single-qubit nodes. We use this device to demonstrate reconfigurable controlled-Z gates across all qubit pairs, with a benchmarked average fidelity of $96.00\%\pm0.08\%$ and best fidelity of $97.14\%\pm0.07\%$, limited mainly by dephasing in the qubits. We also generate multi-qubit entanglement, distributed across the separate modules, demonstrating GHZ-3 and GHZ-4 states with fidelities of $88.15\%\pm0.24\%$ and $75.18\%\pm0.11\%$, respectively. This approach promises efficient scaling to larger-scale quantum circuits, and offers a pathway for implementing quantum algorithms and error correction schemes that benefit from enhanced qubit connectivity.
\end{abstract}

%Use showkeys class option if keyword display desired
\maketitle
%\tableofcontents
\section{Introduction}
%\protect
%\\ %The line
%break was forced \lowercase{via} 
%\textbackslash
%\textbackslash
Quantum computing is the only computational model that can violate the extended Church-Turing thesis, thereby promising exponential speedup on certain problems. Achieving sufficient fidelity for long calculations however requires error correction protocols, that in turn require large numbers of physical qubits, making the construction of quantum computing hardware quite challenging \cite{Fowler2012, Beverland2022, Bravyi2024}. Superconducting qubits provide a promising platform due to their fast operating speeds and ease of fabrication, using which quantum processors have been demonstrated with increasing complexity and fidelity \cite{Wu2021, Krinner2022, Kim2023, Acharya2023}. However, scaling to larger numbers of qubits introduces challenges in fabrication yield, frequency collisions \cite{Hertzberg2021}, and chip-scale correlated errors  \cite{Vepsaelaeinen2020, Cardani2021, Wilen2021, McEwen2021, Harrington2024, Li2024}, among others. A modular approach, including high-fidelity communication channels between separate modules, is one way to sidestep some of these challenges \cite{Gottesman1999, Jiang2007, Xu2022, Smith2022, Bravyi2022}. Recent experimental efforts exploring such concepts include connecting qubit processors with coaxial cables \cite{CampagneIbarcq2018, Leung2019, Zhong2021, Burkhart2021, Yan2022, Storz2023, Renger2023} and coupling modules together using flip-chip techniques \cite{Gold2021, Field2024}. Nevertheless, realizing a deterministic, reconfigurable, and reliable many-to-many connectivity remains an outstanding challenge, compounded by the typical planar structure of most superconducting qubit circuits, with a quadratic overhead in scaling up qubit couplers.

Superconducting qubit interconnects are typically limited to nearest-neighbor couplings on a planar surface, or just a few degrees of enhancement in graph thickness,
% \ANC{what does this mean}\XW{This is a graph theory terminology, meaning the number of planar graph (no edge intersections) into which the original graph edges can be partitioned.}\HY{Can treat the graph thickness as how many physical layers of qubit we need}
even when using multi-layer packaging that combines flip-chip assembly with through-substrate vias \cite{Mallek2021}. More flexible coupling schemes supporting higher connectivity would enable alternative quantum information processing architectures. For example, measurement-based quantum computing requires the generation of high-dimensional entangled states. This relatively unexplored computing paradigm would be facilitated by incorporating many-to-many couplers, obviating the need to transfer entanglement serially through multiple qubits, in which errors can easily propagate across the circuit \cite{Briegel2009}. In addition, quantum error correction codes that leverage higher connectivity between physical qubits promise higher thresholds and lower overheads, possibly outperforming surface code architectures \cite{Xu2024, Bravyi2024}. Analog quantum computing schemes \cite{Das2008, Geller2015, Bernien2017, Song2019, Lu2023} would also benefit from higher connectivity, even with pre-threshold qubits.

\begin{figure*}[t]
    \centering
    \includegraphics{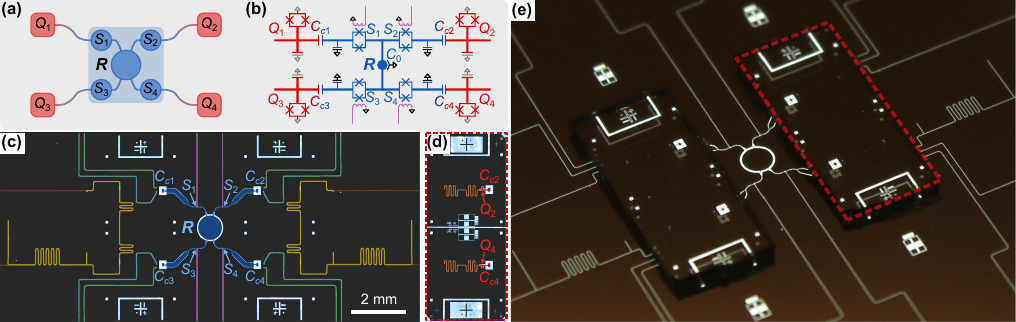}
    \caption{(a) Schematic of the processor: four qubits $Q_i$ (red) are connected to a router $R$ (blue shaded area) consisting of a central capacitor linking four switches $S_i$. (b) Circuit diagram: four Xmon qubits (red) are paired on separate daughterboard substrates and capacitively coupled ($C_{ci}$) to the router (blue), which is a large capacitor ($C_0$) with switches (SQUID loops) attached to each branch, located on the central motherboard. Four flux lines (plum) tune the flux in the router SQUID loops. Daughterboard grounds are grey and motherboard ground is black. Readout circuitry not shown; see Appendix~\ref{sec:measurement_setup}. (c) Layout for the motherboard, including router elements and all control and readout wiring. Switch locations are marked $S_j$; coupling capacitors are marked $C_{ci}$, corresponding to the blue portion of $C_{ci}$ in panel (b).  Each qubit is controlled by a flux line (green) and a drive line (cyan). The readout resonators on the daughterboards (see panel (d)) are pair-wise air-gap inductively coupled to a two-stage Purcell filter (yellow) on the motherboard \cite{Yan2023}, connected to the readout transmission line (brown). (d) Layout for the daughterboard, with two qubits and their dispersive $\lambda/4$ readout resonators (orange); coupling capacitors are marked $C_{ci}$, corresponding to the red portion of $C_{ci}$ in panel (b). (e) Back-lit photo of the assembled device, with two daughterboards placed on top of the motherboard. One daughterboard is dashed-red outlined, corresponding to panel (d).}\label{fig:fig1}
\end{figure*}

One approach to many-to-many connectivity is to link the qubits to a central routing element, similar to a classical ethernet or wireless network. Implementations have included coupling qubits via a common resonator bus \cite{Song2019, Huber2024}, a multi-mode ring resonator \cite{Hazra2021}, or a SNAIL-based quantum state router \cite{Zhou2023}. However, these methods will likely suffer from frequency collisions when scaling to larger qubit numbers, and some require relatively long gate times with complex control protocols, limiting the overall performance.

Here we extend a coupler design for superconducting qubits \cite{Neill2017, Yan2018, Li2020, Collodo2020, Xu2020, Stehlik2021, Sung2021, Sete2021, Goto2022, Zhao2022, Campbell2023, Liang2023, Li2024CZ}, implemented as a type of microwave switch \cite{Pechal2016, Naaman2016, Chang2020}, to realize a tunable multi-mode \cite{McKay2015, Zhang2023} qubit coupler, where the qubits can reside in different chip modules. By pairing each qubit with a SQUID-based variable coupler, we are able to selectively and deterministically link arbitrary pairs of qubits over moderate distances, with negligible interference with other qubits. We implement and then test this processor design by demonstrating a two-qubit controlled-Z (CZ) gate between dynamically selected qubit pairs, and benchmark the gate using cross-entropy benchmarking (XEB) \cite{Boixo2018, Arute2019}. We use these entangling gates to generate multi-qubit entangled states, synthesizing three- and four-qubit GHZ and W states entangled within and across separate modules. 

\section{Experimental design and implementation}
Our experimental design and implementation are shown in Fig.~\ref{fig:fig1}, with the router concept illustrated in panel (a), in which four qubits (red) are coupled to a central routing element (blue shaded area). Each pair of qubits is on a separate modular substrate, with interconnects provided by a supporting motherboard. The corresponding circuit diagram is shown in Fig.~\ref{fig:fig1}(b), in panel (c) the layout for the motherboard, in (d) the layout for one of the two identical modules, and (e) a close-up photograph of the assembled device.

The full circuit comprises four frequency-tunable Xmon superconducting qubits $Q_i$ (red) \cite{Barends2013}, capacitively coupled to four independently-controlled SQUIDs, which operate as router switch elements $S_i$. The switches are galvanically connected to a central capacitor, which together serve as a reconfigurable router $R$ (blue), in this case providing all-to-all connectivity by selectively coupling pairs of qubits through their respective switches. The switch coupling strength is controlled by dynamically tuning the magnetic flux threading each switch's SQUID element. The qubit modules have separate grounds (see Fig.~\ref{fig:fig1}(d)), and are flip-chip assembled on a motherboard where the router and all the control lines are located (see Fig.~\ref{fig:fig1}(c)). To simplify assembly, there are no galvanic connections between the modules and the motherboard, instead using capacitively-coupled microwave grounds. 

The modules and motherboard are fabricated on separate sapphire substrates. An aluminum base layer is first deposited by electron beam evaporation with the base wiring pattern defined by optical lithography followed by reactive plasma etching. The qubit and switch Josephson junctions are lift-off deposited using electron beam lithography and the Dolan bridge method \cite{Dolan1977}. Finally, the modules and motherboard are diced and assembled \cite{Satzinger2019, Conner2021}. Fabrication details can be found in Appendix~\ref{sec:fab}. The assembled device is mounted in a connectorized Al box with Al wire bonds, then mounted and cooled to $10~\mathrm{mK}$ on the mixing chamber stage of a dilution refrigerator. See Appendix~\ref{sec:measurement_setup} for the details of the measurement setup.

The processor operation can be understood as follows: Each SQUID in the router can be treated as a tunable inductor, which reflects (transmits) an incident signal when its inductance is tuned to a relatively large (small) value, representing the `off' (`on') state of the switch. By selectively turning on pairs of switches, we can couple the corresponding qubit pairs, with the other qubits remaining uncoupled. Note that we can in principle couple more than two qubits at the same time, although we have not yet explored this mode of operation.
% \XW{I think I've tried this as another way to generate W-state with improved fidelity, do you think we don't put it in this paper? (which is fine)} \ANC{If you can do this with reasonable fidelity I would say that could be a follow-on paper - 3 and 4 qubit W states in a single process? Are there other interesting things to try?}\XW{Sure. I can also explore more about GHZ with a single process, as well as RL for the application on this and gate synthesis.}

To model the system, we quantize the qubits and switches as coupled Duffing oscillators, using the Hamiltonian 
\begin{equation}\label{eq:duffing}
    \begin{aligned}
        \frac{\mathcal{H}}{\hbar} =& \sum_{i} \left ( \omega_{i} a^{\dagger}_{i} a_{i} + \frac{\eta_{i}}{2} a^{\dagger}_{i} a^{\dagger}_{i} a_{i} a_{i} \right )\\
        &- \sum_{i,j} c_{ij} \left ( a_{i}^{\dagger} - a_{i} \right ) \left( a_{j}^{\dagger} - a_{j} \right),
    \end{aligned}
\end{equation}
where $i$ and $j$ each run over the values $\{Q,S\}\times\{1,2,3,4\}$, with four qubit ($Q_i$) and four switch ($S_i$) modes respectively; $\omega_{i}$ ($\eta_{i}$) represents the frequency (anharmonicity) of mode $i$, while $c_{ij}$ denotes the coupling strength between modes $i$ and $j$. The router is essentially a strongly-coupled, anharmonic multi-mode resonator, formed by the switch SQUID loops and their capacitances. The detuning between the qubit and switch modes is large compared with their coupling strength, including when the switches are on (see Appendix~\ref{sec:effective_freqs}), so we can find an effective Hamiltonian in the qubit subspace using a Schrieffer-Wolff transformation \cite{bravyi_schriefferwolff_2011, Yan2018, Sete2021}. When the qubits are idle, each switch is flux-biased close to its `off' state, with the frequencies of the switch modes sufficiently low that the effective qubit-qubit coupling is negligible. Qubit entangling operations are realized by selectively and dynamically turning on the corresponding switches, which increases each tuned switch's frequency, activating the coupling between the qubits paired with those switches. Single-qubit operations can be performed whenever the paired switch is at its idle (`off') frequency. More details on the qubit-qubit effective coupling $g$ and parasitic qubit $ZZ$ coupling $\zeta$ appear in Appendices~\ref{sec:models} and~\ref{sec:numerical}.

\section{Device operation}
We first demonstrate the functionality and operating protocols of the switches, which involves measuring deterministic population transfers between pairs of qubits, mediated by the switches. This is followed by benchmarking entangling gates between arbitrary qubit pairs.

\begin{figure}[tb]
    \centering
    \includegraphics{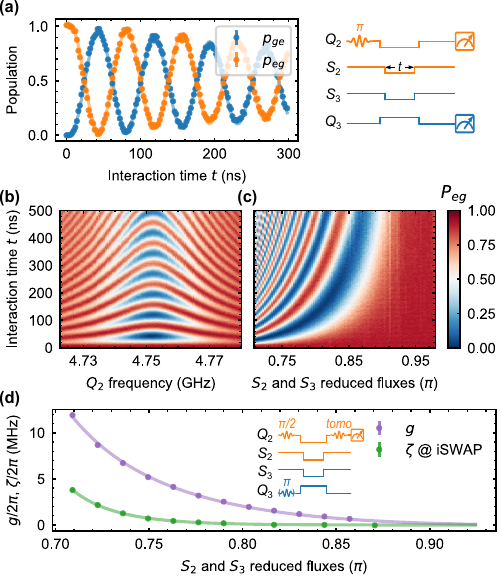}
    \caption{Switch characterization. (a) Rabi swaps between $Q_{2}$ and $Q_{3}$ with the pulse sequence shown on the right, with switch pulses shorter than qubit flux pulses. By (b) sweeping the qubit frequency and (c) varying the external flux applied to each switch (`on' to left, `off' to right), we generate chevron and tunable coupling patterns, respectively. (d) Effective coupling strength $g$ (purple), and parasitic $ZZ$ interaction strength $\zeta$ (green), between $Q_2$ and $Q_3$, as a function of the switch fluxes, sweeping the switch flux from `on' (left) to `off' (right). The pulse sequence for measuring $\zeta$ is inset.} \label{fig:fig2}
\end{figure}

\subsection{Switch calibration}
To benchmark the performance of the switches, shown in Fig.~\ref{fig:fig2}, we first characterize Rabi swaps between on-resonance qubits, turning on the qubit-qubit coupling by flux-biasing the switch pairs simultaneously. In Fig.~\ref{fig:fig2}(a), we excite $Q_{2}$ with all switches off, then bring $Q_{2}$ and $Q_{3}$ on resonance by applying flattop qubit flux pulses. In the meantime, we apply flattop flux pulses to each switch, frequency-tuning $S_{2}$ ($S_{3}$) closer to $Q_{2}$ ($Q_{3}$) to mediate the population transfer. The panel displays the resulting on-resonance swap pattern, where the amplitude decay is dominated by pure dephasing of the qubits. We intentionally place the switch pulses inside the qubit pulses, with a short buffer time, to accommodate the non-zero rise and fall times of the tuning pulses.

Using the same pulse sequence as for panel (a), but now sweeping $Q_{2}$'s frequency, we generate a typical chevron pattern, illustrated in Fig.~\ref{fig:fig2}(b). We similarly demonstrate control of the effective qubit-qubit coupling by simultaneously changing the external fluxes of both $S_{2}$ and $S_{3}$, varying the swap rate between $Q_{2}$ and $Q_{3}$, as shown in Fig.~\ref{fig:fig2}(c). By fitting the swap period, we extract the effective coupling strength $g/2\pi$ as a function of the switch fluxes, shown in Fig.~\ref{fig:fig2}(d) (purple dots). 

In addition, we measure the parasitic $ZZ$ interaction strength $\zeta$ between the on-resonance qubits by performing a cross-Ramsey-type measurement \cite{Sung2021}. The measurement pulse sequence is shown inset in Fig.~\ref{fig:fig2}(d): We prepare $Q_{2}$ in a Bloch sphere equator state with a $\pi/2$ pulse, and simultaneously excite $Q_{3}$ to its $\ket{e}$ state, then turn on the $Q_2-Q_3$ coupling via $S_2$ and $S_3$, resulting in a conditional phase accumulation in $Q_2$. Following two $Q_2-Q_3$ iSWAPs, $Q_{2}$ returns to the equator. The overall accumulated phase is measured by performing state tomography of $Q_2$, from which we extract the $ZZ$ interaction strength $\zeta$. The measured results are shown in Fig.~\ref{fig:fig2}(d) (green dots). In general, $\zeta$ is proportional to $g$, but about one order of magnitude smaller, as a function of the external flux.
% \ANC{what about their proportionality as a function of pulse amplitude?}\XW{$\zeta$ should be proportional to $g$}
The unwanted $ZZ$ coupling during iSWAPs can be further suppressed by engineering the qubit anharmonicities \cite{Zhao2020, Ku2020} or the switch frequencies \cite{Sung2021}, while the CZ gate can circumvent this coupling by adjusting the control phase. At idle points when the switches are `off', the residual $ZZ$ coupling is negligible, which is essential for high-fidelity single-qubit gates and readout. Numerical fits to both the coupling strength and $ZZ$ coupling (solid lines in Fig.~\ref{fig:fig2}(d)) show good agreement with the experimental data (see Appendix~\ref{sec:numerical}).

\begin{figure}[tb]
    \centering
    \includegraphics{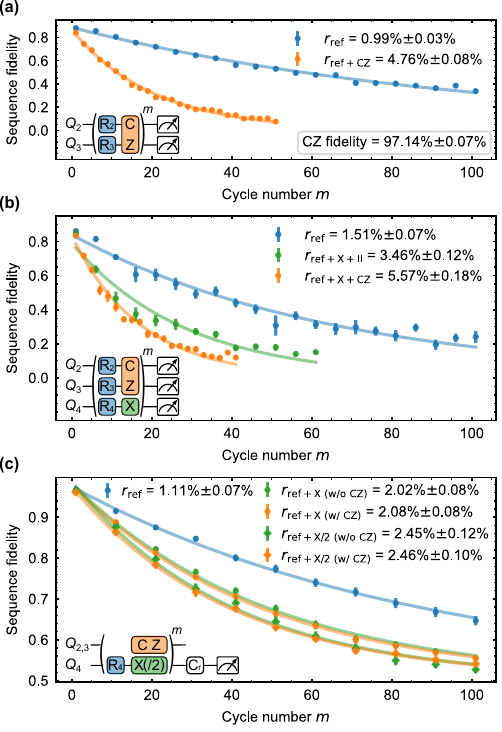}
    \caption{Benchmarking the CZ gate. (a) Cross-entropy benchmarking (XEB), measured by interleaving the CZ gate with random single-qubit gates $R_i$, yielding a CZ fidelity of $97.14\%\pm0.07\%$. (b) Performing XEB as in panel (a) while simultaneously performing interleaved single-qubit gates on qubit $Q_4$. (c) Randomized benchmarking (RB) of a $Q_4$ single-qubit gate with and without a simultaneous $Q_2-Q_3$ CZ gate. Error bars and uncertainties of gate errors are generated by bootstrapping the experimental data \cite{Barends2014}, displayed with $\pm1$ standard deviation.\label{fig:fig3}}
\end{figure}

\subsection{CZ gate benchmarking}
When the coupling between $Q_2$ and $Q_3$ is turned on, we implement a CZ gate by bringing $\ket{Q_2Q_3}=\ket{ee}$ on resonance with $\ket{gf}$, for a duration $\sqrt{2}\pi/2g=43.6~\mathrm{ns}$ \cite{Barends2014}. We benchmark this CZ gate using the cross-entropy benchmarking (XEB) protocol \cite{Boixo2018, Arute2019}, which measures the average overlap between the ideal and the actual probability distributions. To this end, in Fig.~\ref{fig:fig3}(a), we first benchmark single-qubit reference gates $R_i$ randomly selected from the set $\{\mathrm{I}, \mathrm{X}, \mathrm{Y}, \mathrm{X/2}, \mathrm{Y/2}, \mathrm{-X/2}, \mathrm{-Y/2}\}$, which yields a reference error rate of $r_{\mathrm{ref}}=0.99\%\pm0.03\%$. We then interleave a CZ gate within each gate cycle, yielding a total XEB error rate of $r_{\mathrm{ref}+\mathrm{CZ}}=4.76\%\pm0.08\%$, which corresponds to a CZ gate fidelity of $\mathcal{F}=97.14\%\pm0.07\%$ (see Appendix~\ref{sec:benchmarking} for more details on the benchmarking methods). Both reference and interleaved gates are illustrated in the inset quantum circuit. Note that the single-qubit phase changes resulting from the active frequency tuning of the qubits as well as the loading effects due to the switches are already included in the ideal unitary evolution matrix, as these phase changes can be readily corrected by applying single qubit rotations or virtual Z gates \cite{McKay2017}.  We use this method to benchmark CZ gates for all six qubit pairs and obtain an average gate fidelity of $96.00\%\pm0.08\%$, with an average gate time of $46.1~\mathrm{ns}$. All the measured results are provided in Appendix~\ref{sec:cz_tuneup}. Variations between different pairs of qubits arise due to differences in qubit coherence times, differing residual pulse distortions, and unwanted two-level system (TLS) defects. Using numerical simulations that take into account the measured qubit parameters (see Appendix~\ref{sec:device_params}), we can estimate the $T_1$ limit of the gate to be above $99\%$, with dominant errors coming from pure dephasing of the qubits, related to their relatively short $T_2$ times. Errors attributed to residual pulse distortions and leakages can be further suppressed by active pulse-level engineering \cite{Baum2021, Sivak2022, Wu2024}. 

\begin{figure*}[tb]
    \centering
    \includegraphics{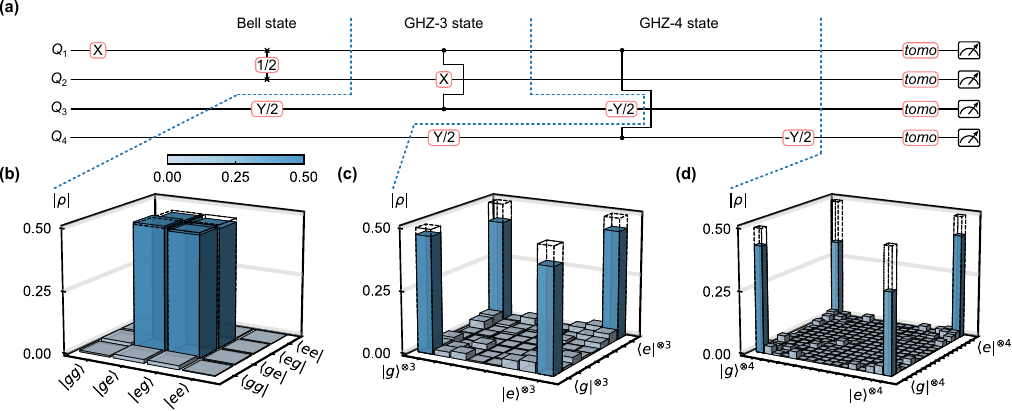}
    \caption{Multi-qubit entanglement generation using one- and two-qubit gates. (a) Quantum circuit for sequentially generating Bell state and $N$-qubit GHZ states ($N = 3, 4$). Dashed blue lines delineate gates used to generate states in panels (b), (c), and (d). (b) Bell state generated by a $\sqrt{\text{iSWAP}}$ gate, yielding a state fidelity $\mathcal{F} = 98.74\%\pm0.25\%$. (c) Three-qubit GHZ state, generated from the Bell state in (b) followed by a CNOT gate between the control qubit and the third qubit, with a fidelity  $\mathcal{F} = 88.15\%\pm0.24\%$. (d) Four-qubit GHZ state generated with another CNOT gate between the control qubit and the fourth qubit, with a fidelity $\mathcal{F} = 75.18\%\pm0.11\%$. All fidelities are defined as $\mathcal{F} = \mathrm{tr}\left( \rho_{\mathrm{ideal}}\rho \right)$, with error bars ($\pm 1$ standard deviation) extracted by bootstrapping the experimental data. Note that single-qubit gates are applied in parallel with the entangling gates to shorten the sequence length, as illustrated in panel (a). \label{fig:fig4}}
\end{figure*}

We also measure the interference between the qubits involved in the CZ gate and the other idle qubits, by simultaneously performing interleaved benchmarking of the CZ between $Q_2$ and $Q_3$, together with single qubit gates on an idle qubit $Q_4$, as shown in Fig.~\ref{fig:fig3}(b). Following measurement of the reference error rate for the three qubits, we interleave an X gate on $Q_{4}$, yielding an error rate of $r_{\mathrm{ref}+\mathrm{X}+\mathrm{II}}=3.46\%\pm0.12\%$ with $Q_2$ and $Q_3$ idle (represented by identity gates `$\mathrm{II}$'). We then repeat this while adding interleaved CZ gates between $Q_2$ and $Q_3$. This yields a total error rate $r_{\mathrm{ref}+\mathrm{X}+\mathrm{CZ}}=5.57\%\pm0.18\%$, corresponding to an inferred CZ fidelity of $97.58\%\pm0.17\%$ considering the X gate error on $Q_{4}$ (see Appendix~\ref{sec:cz_tuneup}). The inferred fidelity is slightly higher than that obtained in Fig.~\ref{fig:fig3}(a), a difference that may be partly due to a slight drift of both single- and two-qubit gate fidelities between experiments. Overall, this result provides clear evidence that the two-qubit entangling gates are not affected by parallel single-qubit gates.

We further verify the negligible interference in Fig.~\ref{fig:fig3}(c) by benchmarking the fidelity of the $Q_4$ single-qubit X and X/2 gates using the standard randomized benchmarking (RB) method \cite{Barends2013}, and compare the RB results measured with and without parallel CZ gates between $Q_2$ and $Q_3$. We find that adding the additional CZ gate only causes the X (X/2) gate fidelity to drop by $0.03\%\pm 0.08\%$ ($0.01\%\pm 0.09\%$).

We use the same technique to benchmark the fidelity of two-qubit iSWAP gates for all six qubit pairs. See Appendix~\ref{sec:cz_tuneup} and~\ref{sec:iswap_tuneup} for more details of the tune-up procedure as well as the benchmarking and numerical evaluation process of these gates.

\section{Multi-qubit entanglement}
Using the benchmarked one- and two-qubit gates, we generate multi-qubit entangled GHZ states involving qubits on different substrates, by sequentially applying two-qubit entangling gates between pairs of qubits. Using the quantum circuit shown in Fig.~\ref{fig:fig4}(a), we first generate a Bell singlet state $\left( \ket{ge} + \ket{eg} \right )/\sqrt{2}$ between $Q_1$ and $Q_2$ via a $\sqrt{\mathrm{iSWAP}}$ gate, which is realized by half-swapping an excitation from $Q_{1}$ to $Q_{2}$. We measure a state fidelity $\mathcal{F} = \mathrm{tr}\left( \rho_{\text{ideal}}\rho \right) = 98.74\%\pm 0.25\%$ using the density matrix in Fig.~\ref{fig:fig4}(b), obtained from state tomography.

We prepare a three-qubit GHZ state $\left( \ket{ggg} + \ket{eee} \right)/\sqrt{2}$ between qubits $Q_1,~Q_2$ and $Q_3$ from the Bell state by performing an X gate on $Q_{2}$, which turns the Bell singlet state into a Bell triplet state $\left( \ket{gg} + \ket{ee} \right)/\sqrt{2}$, together with a CNOT gate constructed from a CZ gate between $Q_{1}$ and $Q_{3}$ combined with single qubit $\pm$Y/2 gates \cite{Barends2013}, the CNOT conditionally flipping $Q_{3}$ based on $Q_{1}$'s state. We measure a state fidelity $\mathcal{F} = \mathrm{tr}\left ( \rho_{\text{ideal}} \rho \right ) = 88.15\%\pm0.24\%$ from the density matrix shown in Fig.~\ref{fig:fig4}(c). The single-qubit phase changes from the CZ gate have been absorbed into the preceding Y/2 gate.

We finally expand the GHZ state to all four qubits using a second CNOT gate between $Q_{1}$ and $Q_{4}$, resulting in the state $\left( \ket{gggg} + \ket{eeee} \right)/\sqrt{2}$ with a fidelity $\mathcal{F} = \mathrm{tr}\left( \rho_{\text{ideal}}\rho \right) = 75.18\%\pm 0.11\%$, with the density matrix shown in Fig.~\ref{fig:fig4}(d). We note this state involves all four qubits on two separate modular substrates. The fidelities of the different GHZ states are characterized using standard state tomography \cite{Zhong2021}, in which combinations of tomography gates $\in\{$I, X/2, Y/2$\}^{n}$ are applied before state measurements along $z$, and the final density matrices reconstructed using convex optimization constrained to ensure physicality. We determine the uncertainties in the state fidelities by bootstrapping the experimental data \cite{Philips2022}, which involves Monte Carlo resampling from the original dataset with replacement and then calculating the standard deviation of the resulting list of fidelities. We note that the infidelities are dominated by gate errors as well as qubit decoherence during the idle periods, which can likely be improved by finding protocols featuring higher parallelism \cite{Song2019, Lu2022, Warren2023}, as the switches form a coupled multi-mode resonator which in principle can mediate simultaneous multi-qubit interactions.

In addition to the GHZ states, we also generate multi-qubit W states using sequential partial swap operations \cite{Hffner2005}. Moreover, by turning all the switches on at the same time, we are able to realize simultaneous swaps between two different qubit pairs, demonstrating the potential for multi-qubit gates, and further exploiting the intrinsic all-to-all connectivity. More details are provided in Appendix~\ref{sec:W} and~\ref{sec:smtns_swap}.

\section{Discussion and outlook}
In conclusion, we describe an on-chip quantum router providing dynamic interconnects in a 4-qubit quantum circuit, which supports tunable, medium-range, all-to-all connectivity. We experimentally demonstrate the design in a modular architecture, benchmarking one- and two-qubit gates between user-selected qubit pairs, using combinations of these gates to generate multi-qubit entanglement. The performance of the device would be improved by longer qubit coherence times and better pulse engineering \cite{Baum2021, Sivak2022, Wu2024}. The star-pattern architecture in Fig.~\ref{fig:fig_circuit_diagram} is easily scaled to a larger number of nodes, limited by the stray capacitance between neighboring branches and by the length of the branch connections. By connecting selected nodes to additional star-pattern networks, a chained architecture can be expanded essentially without limits. In addition, when more than two switches are turned on simultaneously, multi-qubit entangled states \cite{Song2019, Lu2022} or entangling gates \cite{Glaser2023} can be directly synthesized with this processor design. When the quantum processors are placed further apart, our design can likely be adapted to route itinerant microwave photons \cite{Zhong2019, Qiu2023, Grebel2024}. This circuit architecture can also serve as a testbed for investigating many-body quantum dynamics that rely on high connectivity.

\begin{acknowledgments}
Devices and experiments were supported by the the Army Research Office and Laboratory for Physical Sciences (ARO grant W911NF2310077), and by the Air Force Office of Scientific Research (AFOSR grant FA9550-20-1-0270). Results are in part based on work supported by the U.S. Department of Energy Office of Science National Quantum Information Science Research Centers. This work was partially supported by UChicago's MRSEC (NSF award DMR-2011854) and by the NSF QLCI for HQAN (NSF award 2016136). We made use of the Pritzker Nanofabrication Facility, which receives support from SHyNE, a node of the National Science Foundation's National Nanotechnology Coordinated Infrastructure (NSF Grant No. NNCI ECCS-2025633). The authors declare no competing financial interests. Correspondence and requests for materials should be addressed to A. N. Cleland (anc@uchicago.edu).
\end{acknowledgments}

\appendix
\section{Device Fabrication\label{sec:fab}}

\begin{figure*}[tb]
    \centering
    \includegraphics{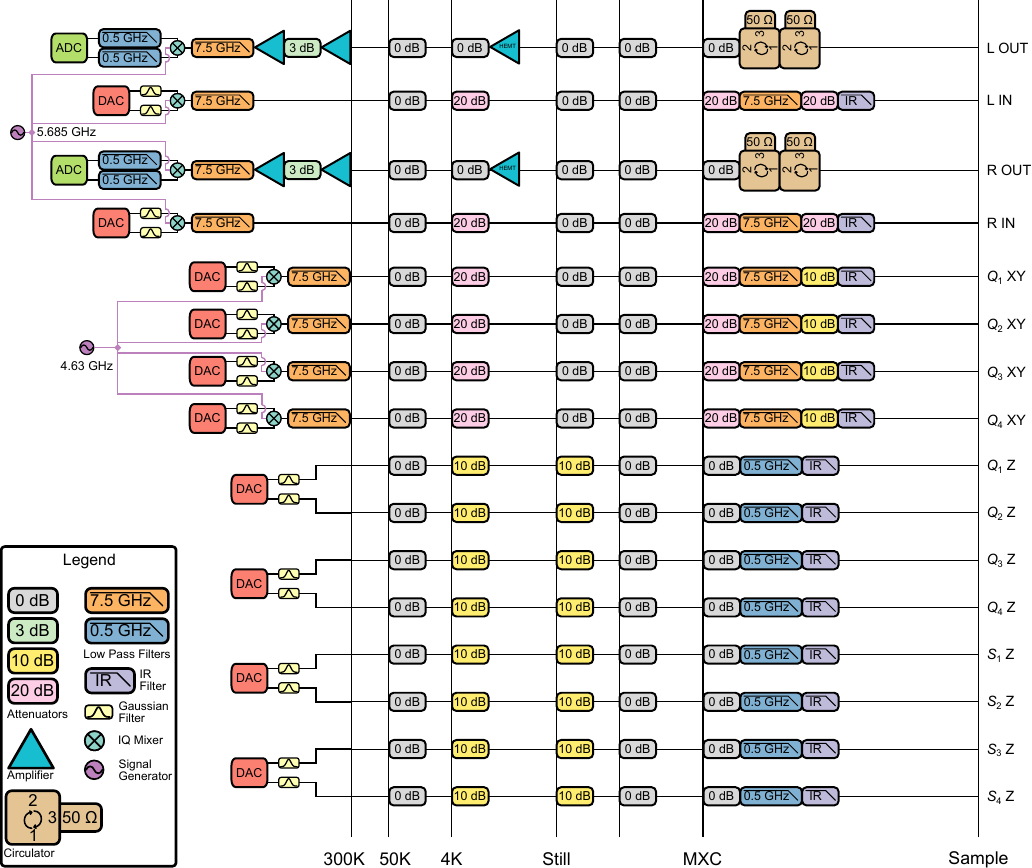}
    \caption{Electronics and wiring diagram for the experiment.\label{fig:fig_wiring}}
\end{figure*}

The fabrication (assembly) process of the motherboard and the daughter modules are adapted from~\cite{Chang2020, Zhong2021} (\cite{Satzinger2019, Conner2021}), with some modifications to improve the robustness of the devices.

A total of 9 motherboards ($20~\mathrm{mm}\times 20~\mathrm{mm}$) and 70 daughter modules ($3~\mathrm{mm}\times 7~\mathrm{mm}$) are fabricated on a 100 mm diameter sapphire wafer, using the following recipe:
\begin{enumerate}
    \item Wafer clean. Before an acetone - isopropyl alcohol - water rinse the wafer is soaked in 50$^{\circ}$C Nanostrip. The wafer is then pre-baked at 200$^\circ$C prior to deposition of Al base layer.
    \item Al base layer. $100$ nm thick Al is deposited on the wafer, then patterned by photolithography followed by a $\mathrm{BCl}_{3}/\mathrm{Cl}_{2}/\mathrm{Ar}$ inductively coupled plasma (ICP) etch to define the base layer circuitry for the qubit and switch capacitors, readout and control lines.
    \item SiO$_2$ layer. $350$ nm SiO$_2$ crossover scaffold is defined and deposited using photolithography followed by electron beam evaporation and liftoff.
    \item Gold layer. $10~\mathrm{nm}/50~\mathrm{nm}$ Ti/Au alignment marks are defined and deposited using photolithography, electron beam evaporation and liftoff.
    \item Wafer quartering. The 100 mm diameter wafer is diced into four quarters, for separate fabrication of Josephson junctions.
    \item Josephson junctions. A PMMA/MAA resist bilayer is patterned with electron beam lithography to define the qubit and switch junctions using a Dolan bridge process \cite{Dolan1977}. No ion milling is involved in this step, so there is no galvanic contact with the base layer metal.
    \item Crossover and bandage layer. $300$ nm thick Al is patterned and liftoff-deposited following an \emph{in~situ} Ar ion mill. This creates superconducting crossovers for all on-chip control wiring, as well as galvanic contacts between the qubit and switch capacitors and the Al base layer.
    \item Spacer layer for the flip-chip. We use SU-8 photodefinable epoxy to define $\sim 6~\mu$m tall studs which are then hard-baked, forming an offset spacer for the subsequent flip-chip process.
    \item Singulation. The quarter-wafer is singulated into motherboard and daughter module dies in preparation for assembly.
    \item Airbridges. A vapor-HF etch removes the $\mathrm{SiO}_{2}$ scaffolds, forming airbridges from the Al crossover metallization on the motherboard substrates.
    \item Flip-chip assembly. One motherboard and two daughter modules are flip-chip assembled using electrically-insulating adhesive.
\end{enumerate}
\section{Measurement setup\label{sec:measurement_setup}}
A color-coded schematic of the room-temperature control electronics and cryogenic wiring is shown in Fig.~\ref{fig:fig_wiring}, similar to that in Ref.~\cite{Zhong2021}. Each DAC and ADC board is dual-channel, providing the functionality for qubit control and measurement respectively, combined with appropriate filtering, attenuation, and amplification along the corresponding lines. A signal generator set to the $4.63~\mathrm{GHz}$ ($5.685~\mathrm{GHz}$) carrier frequency is used for qubit $xy$ control (dispersive readout), with pulse envelopes defined by a DAC dual-channel output signal driving an IQ mixer.
% Qubit and switch flux control (both slow and fast) is implemented with one DAC channel.
For the flux control of all the qubits and switches, instead of using a bias-tee at the $10$ mK stage to combine the slow direct-current (DC) bias from an external voltage source with the fast bias pulse generated by the DAC, we reconfigure the DAC output as well as the cryogenic wiring so that both slow and fast signals are generated and combined within the DAC. Lower cutoff frequency IR filters are incorporated for these control lines for better infrared-photon filtering and thermalization.

\begin{figure}[tb]
    \centering
    \includegraphics{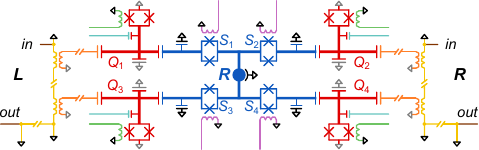}
    \caption{Full circuit diagram of the experimental device.\label{fig:fig_exp_fullcircuit}}
\end{figure}

In Fig.~\ref{fig:fig_exp_fullcircuit} we provide as a supplement to Fig.~\ref{fig:fig1} the full circuit diagram of the device, including all the control lines. Here, green (plum) lines are the flux controls for the tunable Xmon qubits (switches), cyan lines are for qubit drives, while orange and yellow elements represent individual readout resonators and bandpass Purcell filters respectively, which are connected to the readout transmission lines (brown). All the lines are also matched to the sample side in Fig.~\ref{fig:fig_wiring}.
\section{Device characterization\label{sec:device_params}}
Here we present the qubit and switch parameters.
\subsection{Qubit parameters}
We measure basic parameters for all the qubits, which are summarized in Table~\ref{tab:qubit_params}. The maximum frequencies of all qubits are designed to be around $5.4$ GHz, but for some qubits, the actual maximum frequencies are either too close to (or higher than) the readout resonator frequencies, or the qubits couple to two-level states, so we cannot measure the maximum frequencies precisely. The large detunings between the qubit idle frequencies and maximum qubit frequencies limits the pure dephasing time of the qubits~\cite{Yoshihara2006}, which we also observe in devices with tunable qubits on a single planar substrate~(see e.g. \cite{Grebel2024}). This could be allievated by tuning the Josephson junction critical currents \cite{Hertzberg2021, Balaji2024}, or using asymmetric SQUID qubits \cite{Hutchings2017}, where the operating point of the qubit can also be set at or near the minimum frequency. 

To suppress qubit decay due to the Purcell effect while maintaining efficient qubit readout performance, we incorporate a two-stage broadband Purcell filter synthesized using on-chip transmission line segments \cite{Yan2023}. We achieve over $95\%$ readout fidelity without using quantum-limited amplifiers when the corresponding readout resonator frequency falls within the passband of the filter with an appropriate coupling rate. 

Single-qubit gates are benchmarked using randomized benchmarking (RB) (see Appendix~\ref{sec:benchmarking}), where for each qubit we benchmark both X and X/2 gates, yielding the averaged gate fidelity $\mathcal{F}_{\mathrm{SQG}}$.
\begin{table*}[htbp]
 \caption{Qubit parameters. The qubit idle frequency (anharmonicity) is $\omega_{q}$ ($\eta$), $T_{1}$ and $T_{2}^{*}$ are the qubit energy and Ramsey fringe decay times (dominated by dephasing) respectively, and $\mathcal{F}_{\mathrm{SQG}}$ is the average single-qubit gate fidelity measured by randomized benchmarking. $\omega_{rr}$ ($\tau_{rr}$) is the readout frequency (duration), $\chi$ is the dispersive shift, $F_{g}$ ($F_{e}$) is the readout fidelity of the ground state $\ket{g}$ (excited state $\ket{e}$), defined as the measured $\ket{g}$ ($\ket{e}$) probability when the qubit is initialized to the respective state. \label{tab:qubit_params}}
 \centering
 \begin{ruledtabular}
 \bgroup
 \def\arraystretch{1.5}
 \begin{tabular}{ccccc}
   &$Q_{1}$&$Q_{2}$&$Q_{3}$&$Q_{4}$\\
   \hline
   $\omega_{rr}/2\pi$&5.7649 GHz&5.7008 GHz&5.6245 GHz&5.6755 GHz\\
   $\omega_{q}/2\pi$&4.4925 GHz&4.5566 GHz&4.7435 GHz&4.6129 GHz\\
   $\eta/2\pi$&-170 MHz&-181 MHz&-178 MHz&-163 MHz\\
   $\chi/2\pi$&2.1 MHz&2.1 MHz&2.1 MHz&1.3MHz\\
   $\tau_{rr}$&700 ns&800 ns&880 ns&760 ns\\
   $F_{g}$&99.4\%&98.7\%&95.7\%&96.7\%\\
   $F_{e}$&96.7\%&94.1\%&93.1\%&92.9\%\\
   $T_{1}$&29.6 $\mu s$&35.5 $\mu s$&22.1 $\mu s$&14.5 $\mu s$\\
   $T_{2}^{*}$&833 ns&815 ns&1176 ns&818 ns\\
    $\mathcal{F}_{\mathrm{SQG}}$&$99.76\%\pm0.04\%$&$99.64\%\pm0.07\%$&$99.80\%\pm0.03\%$&$99.24\%\pm0.10\%$
 \end{tabular}
 \egroup
 \end{ruledtabular}
\end{table*}
\subsection{Switch response}

\begin{figure}[tb]
    \centering
    \includegraphics{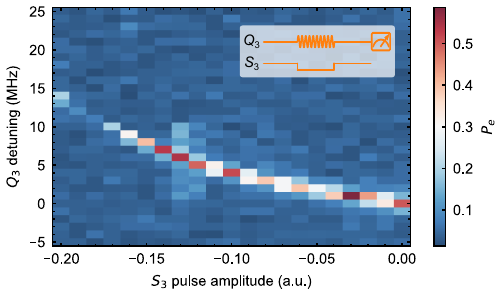}
    \caption{Switch characterization by qubit spectroscopy. We apply a fast flux pulse to bias the switch while pumping the qubit at a frequency detuned from its idle frequency. Pulse sequence is shown inset. \label{fig:fig_switch_response}}
\end{figure}

As will be discussed later in the switch model (Eq.~(\ref{eq:SW})), the frequency of a qubit is sensitive to the flux bias of its respective switch, providing an indirect way to characterize the switch response. An example is shown in Fig.~\ref{fig:fig_switch_response}, where the qubit frequency increases as its switch is flux-biased from its idle point. Since the frequency change is the same order as the effective coupling strength (Eq.~(\ref{eq:SW})), we use this response to identify the proper `off' (right) and `on' (left) points for each switch. The idle frequency of the switch is estimated to be much lower (around $1~\mathrm{GHz}$) than the qubit, so the detuning between their modes is much larger than their coupling strength ($\sim 100-200$~MHz).

\section{Router and qubit model\label{sec:models}}
As has been described previously, our system consists of qubits and a reconfigurable router (switches connected through a central capacitance, each coupled with a corresponding qubit). The switch modes in the router are strongly hybridized and their frequencies with respect to the qubits are readily controlled via locally-applied external magnetic flux, thereby controlling the corresponding qubit-qubit coupling strength.

Here we discuss a Hamiltonian model for the system, including the circuit quantization used to obtain well-defined qubit and switch modes in either two-node or more general $N$-node scenarios, as well as different perturbative methods to calculate effective couplings and parasitic $ZZ$ interactions between qubits mediated by the switch modes \cite{Yan2024}.
\subsection{System quantization\label{sec:sys_quant}}

\begin{figure}[tb]
    \centering
    \includegraphics{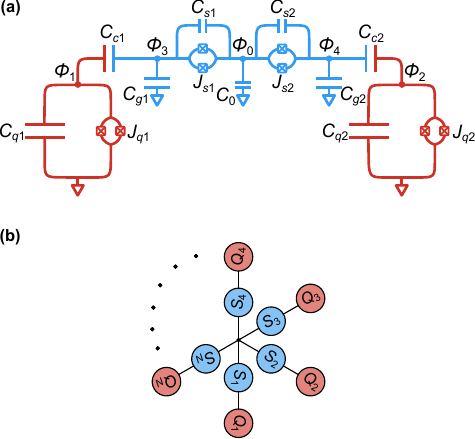}
    \caption{(a) Circuit diagram of the quantum processor comprising two qubits and two switches. (b) Schematic extension to an $N$-node quantum network.} \label{fig:fig_circuit_diagram}
\end{figure}

We first describe the process flow to quantize a circuit consisting of two nodes, and then present the general extension to a quantum network with an arbitrary number of nodes. The circuit diagram for a two-node processor is shown in Fig.~\ref{fig:fig_circuit_diagram}(a), where the red (blue) circuit elements represent the qubit (switch) modes; capacitances $C_{*}$, junctions $J_{*}$ (meaning the effective Josephson inductance of the corresponding SQUID loop), and reduced node fluxes $\phi_{*}$ are labeled separately. Similar to Refs. \cite{Yan2018, Sete2021}, we first write the Lagrangian of the circuit as $\mathcal{L} = \mathcal{T} - \mathcal{U}$,
% \begin{equation}
%     \mathcal{L} = \mathcal{T} - \mathcal{U},
% \end{equation}
where
\begin{equation}
    \begin{aligned}
        \mathcal{T}=\frac{1}{2}\left(\frac{\Phi_0}{2\pi}\right)^2\Big[C_0\dot{\phi}_0^2&+C_{q1}\dot{\phi}_1^2+C_{c1}(\dot{\phi}_3-\dot{\phi}_1)^2\\
        &+C_{g1}\dot{\phi}_3^2+C_{s1}(\dot{\phi}_0-\dot{\phi}_3)^2\\
        &+C_{q2}\dot{\phi}_2^2+C_{c2}(\dot{\phi}_4-\dot{\phi}_2)^2\\
        &+C_{g2}\dot{\phi}_4^2+C_{s2}(\dot{\phi}_0-\dot{\phi}_4)^2\Big]
    \end{aligned}
\end{equation}
is the kinetic energy ($\Phi_{0} = h/2e$ is the magnetic flux quantum), and
\begin{equation}
    \begin{aligned}
    \mathcal{U}=&-E_{J}^{q1}\cos\phi_1-E_{J}^{q2}\cos\phi_2\\
    &-E_{J}^{s1}\cos(\phi_3-\phi_0)-E_{J}^{s2}\cos(\phi_4-\phi_0)
    \end{aligned}
\end{equation}
is the potential energy, with $E_{J} = \hbar I_c/2e = \left(\hbar/2e \right)^{2}/ L_{J}$ the Josephson energy. To diagonalize the nodes in terms of $\mathcal{U}$, we introduce new flux variables given by
\begin{equation}
\begin{aligned}
    \phi_0=\phi_0,~ &\phi_{q1}=\phi_1,~\phi_{s1}=\phi_3-\phi_0,\\
    &\phi_{q2}=\phi_2,~\phi_{s2}=\phi_4-\phi_0.
\end{aligned}
\end{equation}
Then, in terms of these new variables, with $$\bm{\phi}=\left[\phi_0,\phi_{q1},\phi_{s1},\phi_{q2},\phi_{s2}\right]^{\mathrm{T}},$$
$\mathcal{T}$ can be rewritten as
\begin{equation}
    \mathcal{T}=\frac{1}{2}\left(\frac{\Phi_0}{2\pi}\right)^2\dot{\bm{\phi}}^{\mathrm{T}}\bm{C}\dot{\bm{\phi}},
\end{equation}
where the capacitance matrix $\bm{C}$ is
\begin{widetext}
\begin{equation}\label{eq:c_mat}
    \bm{C}=\begin{pmatrix}
        C_0+C_{c1}+C_{g1}+C_{c2}+C_{g2}&-C_{c1}&C_{g1}+C_{c1}&-C_{c2}&C_{g2}+C_{c2}\\
        -C_{c1}&C_{q1}+C_{c1}&-C_{c1}&0&0\\
        C_{g1}+C_{c1}&-C_{c1}&C_{c1}+C_{g1}+C_{s1}&0&0\\
        -C_{c2}&0&0&C_{q2}+C_{c2}&-C_{c2}\\
        C_{g2}+C_{c2}&0&0&-C_{c2}&C_{c2}+C_{g2}+C_{s2}
    \end{pmatrix}.
\end{equation}
\end{widetext}
In order to derive the Hamiltonian, we define the node charges as $\bm{q} = \partial\mathcal{L}/\partial\dot{\bm{\phi}} = \left(\Phi_0/2\pi\right)^2\bm{C}\dot{\bm{\phi}}$, giving
\begin{equation}
    \mathcal{H}=\bm{q}^{\mathrm{T}}\dot{\bm{\phi}}-\mathcal{L}=\frac{1}{2}\left(\frac{2\pi}{\Phi_0}\right)^2\bm{q}^{\mathrm{T}}\bm{C}^{-1}\bm{q}+\mathcal{U}.
\end{equation}
Note that in this expression we will be discarding terms associated with $q_{0}$ and $\phi_{0}$ since $\mathrm{d} q_0/\mathrm{d}t = \partial\mathcal{L}/\partial\phi_{0} = 0$ (Hamilton's equation), which means $q_{0}$ introduces a uniform charge distribution on the metal pads of the qubits or the switches, and thus can be set as arbitrary values without affecting the system dynamics. Applying canonical quantization and defining the Cooper pair number operator as $\bm{n} = \bm{q}/\hbar$, we can express the Hamiltonian as
\begin{equation}\label{eq:H_2node_nphi}
    \begin{aligned}
        \mathcal{H}=~&4E_{C}^{q1}{n}_{q1}^2+4E_{C}^{q2}{n}_{q2}^2+4E_{C}^{s1}{n}_{s1}^2+4E_{C}^{s2}{n}_{s2}^2\\
        +&4E_{C}^{q1,q2}{n}_{q1}{n}_{q2}+4E_{C}^{s1,s2}{n}_{s1}{n}_{s2}\\
        +&4E_{C}^{q1,s1}{n}_{q1}{n}_{s1} +4E_{C}^{q2,s2}{n}_{q2}{n}_{s2}\\
        +&4E_{C}^{q1,s2}{n}_{q1}{n}_{s2} +4E_{C}^{q2,s1}{n}_{q2}{n}_{s1}\\
        -&E_{J}^{q1}\cos{\phi}_{q1}-E_{J}^{q2}\cos{\phi}_{q2}\\
        -&E_{J}^{s1}\cos{\phi}_{s1}-E_{J}^{s2}\cos{\phi}_{s2},
    \end{aligned}
\end{equation}
where $E_C = e^2/2C$ is the charging energy. Specifically, we have $E_C^k = \left(e^2/2\right) \cdot \left[ \bm{C}^{-1} \right]_{kk}$ and $E_C^{kl} = \left(e^2/2\right) \left( \left[
\bm{C}^{-1} \right]_{kl} + \left[ \bm{C}^{-1} \right]_{lk} \right)$, and we have the commutation relations $\comm{{\phi}_{k}}{{n}_{l}}=i\delta_{kl}$, both with $k,l\in\{q_1,q_2,s_1,s_2\}$. We then introduce the photon creation and annihilation operators such that
\begin{equation}
\begin{aligned}
    {n}_k&=\frac{i}{\sqrt{2}}\left(\frac{E_{J}^{k}}{8E_{C}^{k}}\right)^{\frac{1}{4}}\left(a_k^\dagger-a_k\right),\\
    {\phi}_k&=\frac{1}{\sqrt{2}}\left(\frac{8E_{C}^{k}}{E_{J}^{k}}\right)^{\frac{1}{4}}\left(a_k^\dagger+a_k\right),
\end{aligned}
\end{equation}
using which, in the transmon limit $E_{J}\gg E_{C}$, Eq.~(\ref{eq:H_2node_nphi}) can be rewritten in the form of coupled Duffing oscillators,
\begin{equation}\label{eq:H_2node_a}
\begin{aligned}
    \mathcal{H}&=\sum_k\left(\omega_ka^\dagger_ka_k+\frac{\eta_k}{2}a^\dagger_ka^\dagger_ka_ka_k\right)\\
    &-\sum_{k,l}c_{kl}\left(a_{k}^\dagger-a_{k}\right)\left(a_{l}^\dagger-a_{l}\right),
\end{aligned}
\end{equation}
with mode frequencies $\omega_{k}$, anharmonicities $\eta_{k}$, and inter-mode transverse couplings $c_{kl}$ defined as
\begin{align}
    \omega_k&=\sqrt{8E_{J}^{k}E_{C}^{k}}-E_{C}^{k},\label{eq:qubit freq}\\
    \eta_k&=-E_{C}^{k},\label{eq:qubit anh}\\
    c_{kl}&=\frac{E_{C}^{kl}}{\sqrt{2}}\left(\frac{E_{J}^{k}}{E_{C}^{k}}\frac{E_{J}^{l}}{E_{C}^{l}}\right)^{\frac{1}{4}}.\label{eq:coupling}
\end{align}
where we have set $\hbar=1$. Note that the Josephson energies $E_{J}$ are dependent on the external magnetic fluxes threading the dc SQUIDs, given by
\begin{equation}
    E_{J}=E_{J0}\abs{\cos\left( \pi\Phi_{\mathrm{ext}}/\Phi_{0} \right)}=E_{J0}\abs{\cos\left( \phi_{\mathrm{ext}}/2 \right)}
\end{equation}
for symmetric dc SQUIDs, where $\phi_{\mathrm{ext}}=2\pi\Phi_{\mathrm{ext}}/\Phi_0$ is the reduced external flux. The mode frequencies $\omega_{k}$ and couplings $c_{kl}$ therefore vary with $\Phi_{\mathrm{ext}}$.

\subsection{Schrieffer-Wolff transformation\label{sec:SW}}
In order to obtain effective couplings between the qubit modes mediated by the switch modes, we block-diagonalize Eq.~(\ref{eq:H_2node_a}) so that only intra-qubit or intra-switch couplings remain. A systematic approach is to perform a Schrieffer-Wolff transformation \cite{bravyi_schriefferwolff_2011} to eliminate the couplings between the qubit and switch modes perturbatively. Because of the large bare intra-switch coupling (approaching the order of qubit-switch detuning), the standard transformation $\mathcal{U} = e^{\mathcal{S}}$, simply based on both rapid counter-rotating wave terms and slow rotating wave terms \cite{Yan2018}, does not apply. Instead we take $\mathcal{S}$ in the form
\begin{equation}
\begin{aligned}
        \mathcal{S}=\sum_{m,n}~&x_{mn}\left(a_{qm}^\dagger a_{sn}-a_{qm}a_{sn}^\dagger\right)\\\quad+&y_{mn}\left(a_{qm}^\dagger a_{sn}^\dagger-a_{qm}a_{sn}\right),
\end{aligned}
\end{equation}
where $x_{mn}$ and $y_{mn}$ are coefficients to be determined, with mode indices $m$ and $n$. Note that Eq.~(\ref{eq:H_2node_a}) can be grouped as
\begin{equation}
    \mathcal{H}=\mathcal{H}_{q}+\mathcal{H}_{s}+\mathcal{H}_{qq} + \mathcal{H}_{ss} + \mathcal{H}_{qs},
\end{equation}
so to the second order, $\mathcal{S}$ is determined by solving
\begin{equation}
    \mathcal{H}_{qs} + \comm{\mathcal{S}}{\mathcal{H}_{q} + \mathcal{H}_{s} + \mathcal{H}_{qq} + \mathcal{H}_{ss}}=0.
\end{equation}
In the two-node scenario, we arrive at a linear system of equations,
\begin{widetext}
\begin{equation}\label{eq:xy_coeff}
    \begin{pmatrix}
        0&0&-c_{s1s2}&c_{q1q2}&-\Sigma_{q1s1}&0&-c_{s1s2}&-c_{q1q2}\\
        0&0&c_{q1q2}&-c_{s1s2}&0&-\Sigma_{q2s2}&-c_{q1q2}&-c_{s1s2}\\
        -c_{s1s2}&c_{q1q2}&0&0&-c_{s1s2}&-c_{q1q2}&-\Sigma_{q1s2}&0\\
        c_{q1q2}&-c_{s1s2}&0&0&-c_{q1q2}&-c_{s1s2}&0&-\Sigma_{q2s1}\\
        -\Delta_{q1s1}&0&c_{s1s2}&-c_{q1q2}&0&0&c_{s1s2}&c_{q1q2}\\
        0&-\Delta_{q2s2}&-c_{q1q2}&c_{s1s2}&0&0&c_{q1q2}&c_{s1s2}\\
        c_{s1s2}&-c_{q1q2}&-\Delta_{q1s2}&0&c_{s1s2}&c_{q1q2}&0&0\\
        -c_{q1q2}&c_{s1s2}&0&-\Delta_{q2s1}&c_{q1q2}&c_{s1s2}&0&0
    \end{pmatrix}\begin{pmatrix}
        x_{11}\\x_{22}\\x_{12}\\x_{21}\\y_{11}\\y_{22}\\y_{12}\\y_{21}
    \end{pmatrix}=\begin{pmatrix}
        c_{q1s1}\\c_{q2s2}\\c_{q1s2}\\c_{q2s1}\\-c_{q1s1}\\-c_{q2s2}\\-c_{q1s2}\\-c_{q2s1}
    \end{pmatrix},
\end{equation}
\end{widetext}
where $\Sigma_{qmsn}=\omega_{qm}+\omega_{sn}$ and $\Delta_{qmsn}=\omega_{qm}-\omega_{sn}$, and again $m$ and $n$ are the mode indices. Here we neglect terms associated with the Kerr nonlinearities $\eta_{k}$, which is a good approximation as long as the switch populations remain mostly in the ground state \cite{Yan2018}. The coefficients can be obtained by inverting the matrix in Eq.~(\ref{eq:xy_coeff}), although it is worth mentioning that in the case of small intra-mode couplings $\left( c_{q1q2}, c_{s1s2}\ll \abs{\Delta_{qmsn}} \right)$, the coefficients are $x_{mn}=c_{qmsn}/\Delta_{qmsn}$ and $y_{mn}=-c_{qmsn}/\Sigma_{qmsn}$, which are exactly the slow rotating wave and fast counter-rotating wave terms. 

The transformed Hamiltonian, to second order, can be obtained by using $\mathcal{S}$ in the transformation
\begin{equation}\label{eq:SW_2nd_order}
    \begin{aligned}
    \mathcal{H}^{\prime}={}&e^{\mathcal{S}}\mathcal{H}e^{-\mathcal{S}}=\mathcal{H}^{\prime}_{q}+\mathcal{H}^{\prime}_{s}+\mathcal{H}^{\prime}_{qq} + \mathcal{H}^{\prime}_{ss}\\
    ={}&\mathcal{H}_{q}+\mathcal{H}_{s}+\mathcal{H}_{qq} + \mathcal{H}_{ss}+\frac{1}{2}\comm{\mathcal{S}}{\mathcal{H}_{qs}},
    \end{aligned}
\end{equation}
from which we are able to get the effective mode frequencies as well as the coupling strengths. The effective coupling strength between the qubits becomes
\begin{equation}
\begin{aligned}
    g_{q1q2}={}&c_{q1q2}+\\
    &\frac{1}{2}
    \begin{alignedat}[t]{1}
    \Big[&(x_{11}+y_{11})c_{q2s1}+(x_{22}+y_{22})c_{q1s2}+\\
    &(x_{12}+y_{12})c_{q2s2}+(x_{21}+y_{21})c_{q1s1}\Big],
    \end{alignedat}
\end{aligned}
\end{equation}
indicating a mixing effect from all intra- and inter-mode couplings at certain mode detunings.

The model involving an arbitrary number of nodes (see Fig.~\ref{fig:fig_circuit_diagram}(b)) can be developed in a similar way, where now in Eq.~(\ref{eq:H_2node_a}), both indices $k,l\in\{q,s\}\times\{m=1,2,..., N\}$, and in turn the linear system in Eq.~(\ref{eq:xy_coeff}) for solving the second-order Schrieffer-Wolff transformation can be generalized as
\begin{align}
    -&\Delta_{qmsn}x_{mn}-\sum_{t\neq m}c_{qmqt}x_{tn}+\sum_{t\neq n}c_{stsn}x_{mt}\nonumber\\
    +&\sum_{t\neq m}c_{qmqt}y_{tn}+\sum_{t\neq n}c_{stsn}y_{mt}+c_{qmsn}=0,\\
    -&\Sigma_{qmsn}y_{mn}+\sum_{t\neq m}c_{qmqt}x_{tn}-\sum_{t\neq n}c_{stsn}x_{mt}\nonumber\\
    -&\sum_{t\neq m}c_{qmqt}y_{tn}-\sum_{t\neq n}c_{stsn}y_{mt}-c_{qmsn}=0.
\end{align}
Here both $m,n$ range from $1$ to $N$, representing a total of $2N^{2}$ equations. After the transformation, the coupled multi-mode system takes on the block-diagonalized form (Eq.~(\ref{eq:SW_2nd_order})), with effective parameters
\begin{align}\label{eq:SW}
\bm{\omega}_{q}^{\prime} = {}&\bm{\omega}_{q} + \left[ \left( \bm{x} + \bm{y} \right)\bm{c}_{qs}^{\mathrm{T}} \right]_{\text{diag}},\\
\bm{\omega}_{s}^{\prime} = {}&\bm{\omega}_{s} - \left[ \left( \bm{x} - \bm{y} \right)^{\mathrm{T}}\bm{c}_{qs} \right]_{\text{diag}},\\
\bm{g}_{qq} = {}&\bm{c}_{qq} + \frac{1}{2}\left( \bm{x} + \bm{y} \right)\bm{c}_{qs}^{\mathrm{T}} + \frac{1}{2}\bm{c}_{qs}\left( \bm{x} + \bm{y} \right)^{\mathrm{T}},\\
\bm{g}_{ss} = {}&\bm{c}_{ss} - \frac{1}{2}\left( \bm{x} - \bm{y} \right)^{\mathrm{T}}\bm{c}_{qs} - \frac{1}{2}\bm{c}_{qs}^{\mathrm{T}}\left( \bm{x} - \bm{y} \right),
\end{align}
where the effective frequencies $\bm{\omega}_{q}^{\prime}$ and $\bm{\omega}_{s}^{\prime}$ (effective couplings $\bm{g}_{qq}$ and $\bm{g}_{ss}$) are written in vector (matrix) form, and the coefficients $\bm{x}$ and $\bm{y}$ are reshaped as $N\times N$ matrices.

Other than the effective coupling strength $g$, we can in principle also calculate the parasitic $ZZ$ interaction strength $\zeta$ by going to fourth order~\cite{Yan2018, Sung2021, Sete2021}. However, this becomes much more difficult here, as strongly coupled switch modes are involved. We instead calculate $\zeta$ using the impedance response method discussed next.

\subsection{Impedance response formulas (Z-method)\label{sec:impedance_response}}

\begin{figure}[tb]
    \centering
    \includegraphics{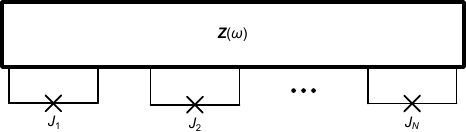}
    \caption{Circuit schematic for $N$ qubits coupled via an $N$-port network \cite{Solgun2019}.} \label{fig:fig_Z_method_Nport}
\end{figure}

An efficient perturbative method to compute $g$ and $\zeta$ for a multi-qubit system is introduced in Refs.~\cite{Solgun2019, Solgun2022}. Here we briefly review their main results, showing the compatibility with our model. As illustrated in Fig.~\ref{fig:fig_Z_method_Nport}, any system involving $N$ transmon qubits can be represented by an $N$-port network where all the qubits are coupled at the ports, with their capacitors absorbed into the network. Similar to the process in Appendix~\ref{sec:sys_quant}, we can obtain the qubit frequencies $\omega_k$ and the anharmonicities $\eta_k$ by solving the capacitance matrix $\bm{C}$. Note that Ref.~\cite{Solgun2022} added small corrections to Eqs.~(\ref{eq:qubit freq}) and (\ref{eq:qubit anh}) to make the results more precise.

The effective coupling strength between qubits $k$ and $l$ can be formulated as
\begin{equation}\label{eq:imp_g}
    g_{kl} = -\frac{1}{4}\sqrt{\frac{\omega_k\omega_l}{L_kL_l}}\mathrm{Im}\left[\frac{Z_{kl}(\omega_k)}{\omega_k}+\frac{Z_{kl}(\omega_{l})}{\omega_l}\right].
\end{equation}
For the switch modes in $\bm{Z}(\omega)$ which contain Josephson junctions as nonlinear elements, we can linearize by replacing $L_J$ with $L=1/\omega^2 C=L_J/(1-2E_C/\omega)$, where $\omega$ is the frequency of the corresponding switch mode, as long as these are in the transmon limit $E_J\gg E_C$.

To calculate $\zeta_{kl}$, we need to first satisfy the condition $\abs{g_{kl}} \ll \abs{\omega_k - \omega_l}$, then $\zeta_{kl}$ can be decomposed as $\zeta_{kl} = \zeta_K^{kl} + \zeta_J^{kl}$, where  $\zeta_K^{kl}$ is the direct $ZZ$ coupling and $\zeta_J^{kl}$ is the $ZZ$ coupling due to the exchange coupling. These terms are expressed as
\begin{align}
        \zeta_K^{kl} &= 2\eta_k\left(\frac{\omega_k}{\omega_l}\right)\alpha_{kl}^2+2\eta_l\left(\frac{\omega_l}{\omega_k}\right)\alpha_{lk}^2,\label{eq:imp_zeta_K}\\
        \zeta_J^{kl} &= 2\frac{J_1^2(\eta_l-\Delta_{kl})+J_2^2(\eta_k+\Delta_{kl})}{(\Delta_{kl}+\eta_k)(\Delta_{kl}-\eta_l)},\label{eq:imp_zeta_J}
\end{align}
where $\alpha_{kl}=\sqrt{C_k/L_l}\frac{1}{2\left (\omega_k^2-\omega_l^2\right)}\mathrm{Im}\Big[\left(\omega_k^2-2\omega_l^2\right)Z_{kl}(\omega_l)+\omega_k\omega_l Z_{kl}(\omega_k)\Big]$, $\Delta_{kl} = \omega_{k} - \omega_{l}$, and $J_k$ is the effective coupling strength between the higher excited states, given by
\begin{align}
    J_1&=-\frac{1}{4}\sqrt{\frac{\omega_k\omega_l}{L_kL_l}}\mathrm{Im}\left[\beta_{11}^{kl}\frac{Z_{kl}(\omega_k)}{\omega_k}+\beta_{12}^{kl}\frac{Z_{kl}(\omega_{l})}{\omega_l}\right],\\
    J_2&=-\frac{1}{4}\sqrt{\frac{\omega_k\omega_l}{L_kL_l}}\mathrm{Im}\left[\beta_{21}^{kl}\frac{Z_{kl}(\omega_k)}{\omega_k}+\beta_{22}^{kl}\frac{Z_{kl}(\omega_{l})}{\omega_l}\right],
\end{align}
with the $\beta$ correction coefficients
\begin{align}
    \beta_{11}^{kl} & = 1+2\frac{\omega_k\eta_k}{\omega_k^2-\omega_l^2},\\
        \beta_{12}^{kl} & = 1- 2\frac{\omega_k\eta_k}{\omega_k^2-\omega_l^2} + 4\frac{\eta_k}{\omega_k},\\
    \beta_{21}^{kl} & = 1- 2\frac{\omega_l\eta_l}{\omega_l^2-\omega_k^2} + 4\frac{\eta_l}{\omega_l},\\
        \beta_{22}^{kl} & = 1+2\frac{\omega_l\eta_l}{\omega_l^2-\omega_k^2}.
\end{align}

With the impedance matrix $\bm{Z}(\omega)$, we can calculate $g$ and $\zeta$ using Eqs.~(\ref{eq:imp_g}), (\ref{eq:imp_zeta_K}), and (\ref{eq:imp_zeta_J}).

\section{Numerical simulations\label{sec:numerical}}
Here we present simulation results for $g$ and $\zeta$ based on the models in Appendix~\ref{sec:models}. We also describe the process to numerically simulate the system dynamics, which will be adopted in Appendix~\ref{sec:cz_tuneup} and \ref{sec:iswap_tuneup} for evaluating the performance of entangling gates.

\subsection{Effective coupling\label{sec:geff_simu}}

\begin{figure}[tb]
    \centering
    \includegraphics{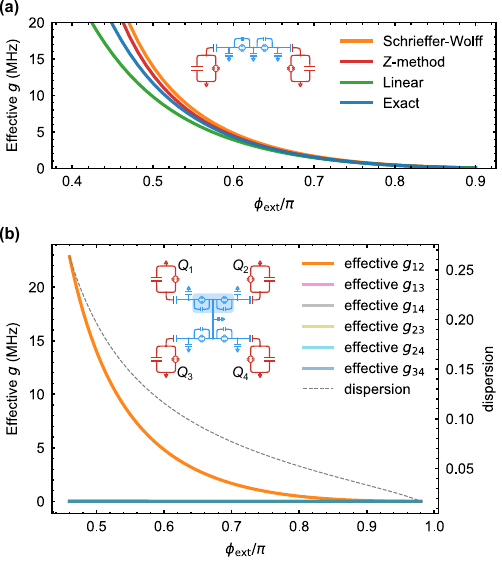}
    \caption{Simulation of switch-mediated effective coupling strength $g$. (a) For a two-node quantum processor, we numerically simulate the effective coupling strength between the two qubits as a function of the reduced external magnetic flux $\phi_{\mathrm{ext}}$ applied to each switch. We compare the results obtained by four different methods and find reasonable agreement. Here `linear' represents the classical method, where all junctions are replaced with linear inductors, while `exact' refers to calculations using the SQcircuit package~\cite{Rajabzadeh2023}. (b) For the four-node processor, we simultaneously turn on the switches paired with $Q_{1}$ and $Q_{2}$ (shaded area), while leaving other switches at their idle points. From the simulation, we see that only the effective coupling between $Q_{1}$ and $Q_{2}$ is switched on, with all the other qubit pairs remaining uncoupled. Here we adopt the Schrieffer-Wolff transformation method for more efficient simulations for the larger number of nodes. We also plot the dispersive coefficient (labeled `dispersion') as a function of the flux, showing it remains in the dispersive limit when the effective coupling strength is a few MHz. The dispersive coefficient is defined as $\mathrm{max}\left\{ \abs{x_{mn}}, \abs{y_{mn}} \right\}$, where both $m, n$ range from $1$ to $N$. \label{fig:fig_geff_simu}}
\end{figure}

We simulate the effective coupling strength $g$ on both two-node and four-node processor models (see Fig.~\ref{fig:fig_geff_simu}), using identical circuit parameters for each node, with $C_q = 100~\mathrm{fF},~C_c = 20~\mathrm{fF},~C_s = 7.5~\mathrm{fF},~C_g = 420~\mathrm{fF},~C_0 = 780~\mathrm{fF},~J_q = 8~\mathrm{nH},~\mathrm{and}~J_s = 4~\mathrm{nH}$. No external fluxes are applied to the qubits, so all the qubits remain close to resonance, taking into account loading from the switches. For the two-node model, we simulate $g$ as a function of the external flux $\phi_{\mathrm{ext}}$ applied to each switch using four different methods, showcasing the high on-off ratio attainable with tunable coupling. For the four-node model, we demonstrate the selectivity of the coupling scheme by turning on two out of four switches, where only the coupling between the corresponding qubits is affected as expected.

\subsection{$ZZ$ interaction}

\begin{figure}[tb]
    \centering
    \includegraphics{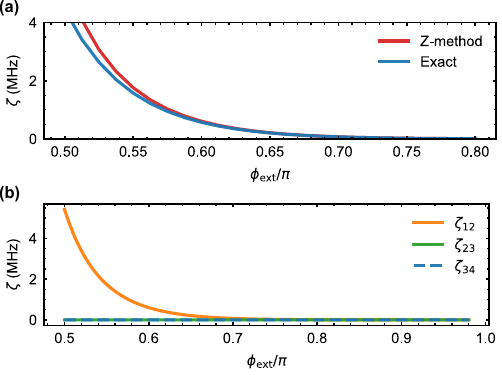}
    \caption{Simulation of switch-mediated $ZZ$ interaction strength $\zeta$. Schematics are the same as the insets in Fig.~\ref{fig:fig_geff_simu}. (a) For the two-node processor, we use both the Z-method and exact diagonalization to compute $\zeta$ as a function of the flux $\phi_{\mathrm{ext}}$ applied to each switch, showing good agreement between the two methods. (b) For the four-node processor, the flux $\phi_{\mathrm{ext}}$ is only applied to $S_1$ and $S_2$, so only $\zeta_{12}$ is affected, as expected.} \label{fig:fig_zeta_simu}
\end{figure}

We also simulate the $ZZ$ interaction strength $\zeta$ on both two-node and four-node processor models (see Fig.~\ref{fig:fig_zeta_simu}), using the same circuit parameters as in Appendix~\ref{sec:geff_simu}, other than biasing some of the qubits to achieve the condition $\abs{g_{kl}} \ll \abs{\omega_k - \omega_l}$. Specifically, in both panels (a) and (b) of Fig.~\ref{fig:fig_zeta_simu}, we set $\left(\omega_{q2}-\omega_{q1}\right)/2\pi = 66~\mathrm{MHz}$, while in (b) we also set $\omega_{q3} = \omega_{q1}$ and $\omega_{q4} = \omega_{q2}$. We demonstrate both the tunability and selectivity of $\zeta$, where one important result is that any idle qubit has negligible residual $ZZ$ interaction with the other qubits, an essential requirement for both high-fidelity gate operations and readout.

\subsection{Effective qubit and switch mode frequencies\label{sec:effective_freqs}}

\begin{figure}[tb]
    \centering
    \includegraphics{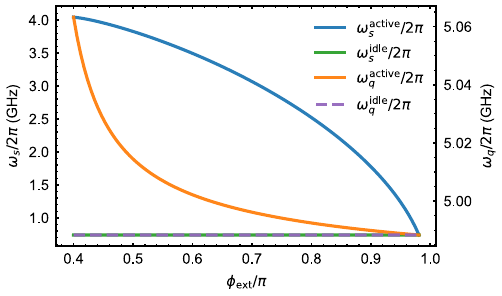}
    \caption{Simulation of effective qubit and switch mode frequencies with respect to the reduced external flux $\phi_{\mathrm{ext}}$ applied to each switch. Here the model is the same as in Fig.~\ref{fig:fig_geff_simu}(b), where we simultaneously turn on $S_1$ and $S_2$ with the same flux bias. Therefore, `active' (`idle') means indices `1, 2' (`3, 4') respectively.} \label{fig:fig_omega_simu}
\end{figure}

We are able to calculate the effective qubit and switch mode frequencies in our model using the Schrieffer-Wolff transformation method, with the results shown in Fig.~\ref{fig:fig_omega_simu}. We see that when the switches are in the `off' state, their idle frequencies are lower than 1 GHz, sufficiently detuned from the qubit modes to leave negligible coupling. When two switches are turned on, their mode frequencies increase, typically to about $1 \sim 2$ GHz from the qubit frequencies, thus mediating an effective qubit-qubit coupling. During the flux tuning process, the frequencies of the active qubits are also perturbed while those of the idle qubits and switches remain unchanged. It is also worth noting that, while not directly measured in the experiment, the switch modes feature around 65 MHz anharmonicities, which are estimated from the circuit parameters in Appendix~\ref{sec:geff_simu}.

\subsection{System dynamics}
Numerical simulations of the circuit dynamics are based on the Hamiltonian in Eq.~(\ref{eq:H_2node_a}). Here we transform to the rotating frame defined by the idle frequencies of each individual mode, including all qubits and switches. Applying the rotating wave approximation (RWA), we arrive at the rotating frame Hamiltonian
\begin{equation}\label{eq:H_rot}
    \begin{aligned} \mathcal{H}_{\mathrm{rot}}&=\sum_k\left(\Delta_ka^\dagger_ka_k+\frac{\eta_k}{2}a^\dagger_ka^\dagger_ka_ka_k\right)\\
        &+\sum_{k,l}c_{kl}\left( e^{i\delta_{0}^{kl}t}a_{k}^{\dagger}a_{l} + e^{-i\delta_{0}^{kl}t}a_{k}a_{l}^{\dagger} \right),
\end{aligned}
\end{equation}
where $\Delta_{k} = \omega_{k} - \omega_{0}^{k}$ is the detuning of mode $k$ with respect to the corresponding idle frequency $\omega_{0}^{k}$, and $\delta_{0}^{kl} = \omega_{0}^{k} - \omega_{0}^{l}$ is the difference between two idle frequencies. $\Delta_{k}$ and $c_{kl}$ become time-dependent as we tune the mode fluxes. In the meantime, we have decoherence in the system, which can be modeled using the jump operators in the Lindblad master equation. This is formulated as
\begin{equation}\label{eq:me}
    \dot{\rho} = -i\comm{\mathcal{H}_{\mathrm{rot}}}{\rho} + \sum\limits_{k}\gamma_{k}\left( \mathcal{L}_{k}\rho\mathcal{L}_{k}^{\dagger} - \frac{1}{2}\acomm{\mathcal{L}_{k}^{\dagger}\mathcal{L}_{k}}{\rho} \right),
\end{equation}
where $\gamma_{k} \sim 1/T_{1}~( 1/T_{\phi} )$ and $\mathcal{L}_{k} = a_{k}~( a_{k}^{\dagger}a_{k} )$ for energy decay (pure dephasing). We also include thermal populations of the initial states of the switches, as their idle frequencies are relatively low.

\section{Gate benchmarking schemes\label{sec:benchmarking}}
We use the randomized benchmarking (RB) (cross-entropy benchmarking (XEB)) procedure for evaluating the single-qubit (two-qubit) gates. Here we briefly review the implementation of both methods.
\subsection{Randomized benchmarking (RB)}
\begin{figure}[tb]
    \centering
    \includegraphics{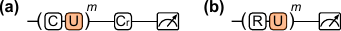}
    \caption{Benchmarking sequence diagram, wherein (a) RB has randomized reference Clifford gates $C$ interleaved with the benchmarking target gate $U$, and a recovery gate $C_{r}$ in the end, and (b) XEB has randomized reference gates $R$ interleaved with the benchmarked gate $U$.} \label{fig:fig_rb_xeb_seq}
\end{figure}
RB refers to a family of benchmarking protocols where random gates are applied to the quantum system to traverse the noise space and thereby obtain an average decay rate for quantum information in the system. Among the various forms of RB, Clifford group-based RB methods stand out as one of the most commonly used prototypes. The RB gate sequence is shown in Fig.~\ref{fig:fig_rb_xeb_seq}(a), where repeated reference and interleaved gates are applied up to some circuit depth $m$, followed by a terminal recovery gate. We extract the sequence fidelity, defined as
\begin{equation}\label{eq:rb_1}
    \mathcal{F}(m) = A \cdot p^m + B,
\end{equation}
where $A$ and $B$ are coefficients related to state preparation and measurement (SPAM) errors, where in the ideal case with no SPAM errors, $A = 1 - 1/d$ and $B = 1/d$, $d$ being the system dimension. For both reference and interleaved benchmarking, we can obtain the decay rates $p_{\mathrm{ref}}$ and $p_{U}$ from Eq.~(\ref{eq:rb_1}), from which the gate fidelity $\mathcal{F}(U)$ is calculated as \cite{Barends2013}
\begin{equation}\label{eq:rb_2}
    \mathcal{F}(U) = 1 - \frac{d-1}{d}\left( 1 - \frac{p_U}{p_{\mathrm{ref}}} \right).
\end{equation}
Note that as we apply a recovery gate at the end of each benchmarking sequence, the benchmarking target gate $U$ is usually limited to the Clifford set.
\subsection{Cross-entropy benchmarking (XEB)}
XEB is a powerful tool designed specifically for benchmarking large quantum processors \cite{Boixo2018, Arute2019}, where the method is to measure the fidelity of certain quantum processes by comparing the terminal experimental and the ideal probability distributions using the cross-entropy metric. One common way to formulate XEB is by measuring $\mathcal{O} = \sum_{x}p_{\mathrm{ideal}}\dyad{x}$ at a certain circuit depth $m$. Assuming a depolarization noise model $\rho_{\mathrm{exp}} = \mathcal{F}(m) \rho_{\mathrm{ideal}} + \left( 
1 - \mathcal{F}(m) \right) \mathds{1} / d$ and observing $\mathcal{O}$ on both sides, we arrive at
\begin{equation}
\begin{aligned}
    \sum\limits_{x}p_{\mathrm{exp}}(x)p_{\mathrm{ideal}}(x) = {}&\mathcal{F}(m)\sum\limits_{x}p_{\mathrm{ideal}}^{2}(x)\\
    &+\left( 1 - \mathcal{F}(m) \right) / d.
\end{aligned}
\end{equation}
The XEB fidelity $\mathcal{F}(m)$ is evaluated for each circuit depth $m$ by executing different random circuits. The decay rate $p$ is extracted by fitting the fidelity to Eq.~(\ref{eq:rb_1}), where in the ideal case (no SPAM errors) we have $A = 1$ and $B = 0$. The gate fidelity $\mathcal{F}(U)$ is then calculated from Eq.~(\ref{eq:rb_2}), where the error rate $r$ in Fig.~\ref{fig:fig3} is defined as $r=1-p$. Here the target gate is generally not restricted to the Clifford group, which is an advantage of XEB over RB, especially when the system dimension is large.

\begin{figure}[tb]
    \centering
    \includegraphics{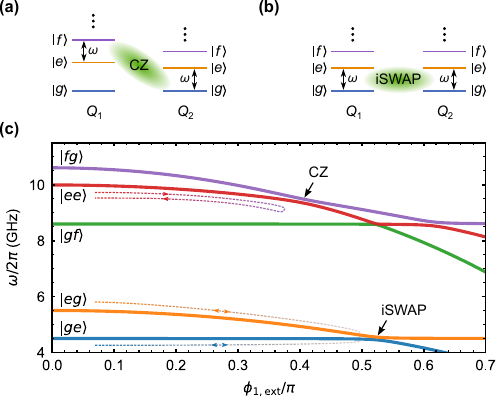}
    \caption{Entangling gates implementation. (a) The CZ gate is realized by resonantly interacting $\ket{ee}$ with $\ket{fg}$ when $Q_{1}$ is blue-detuned with its anharmonicity relative to $Q_{2}$. Such transition is activated for a round trip when all the population reverts back to $\ket{ee}$ but with an accompanying $\pi$ phase accumulation. It also works when $Q_{1}$ is red-detuned with $Q_{2}$'s anharmonicity, where instead $\ket{ee}$ interacts with $\ket{gf}$. (b) The iSWAP gate is realized by simply tuning both qubits on resonance and waiting for a duration when $\ket{ge}$ ($\ket{eg}$) fully evolves to $\ket{eg}$ ($\ket{ge}$). (c) Energy diagram of two coupled qubits as a function of the reduced external flux applied to $Q_{1}$. Here the CZ gate in (a) is applied at the avoided crossing point between $\ket{ee}$ and $\ket{fg}$, while the iSWAP gate in (b) is applied at the avoided crossing point between $\ket{ge}$ and $\ket{eg}$. Dashed lines with arrows indicate the typical trajectories for implementing the gates. Note that although the iSWAP gate inevitably passes the CZ operation point, our buffer time between the qubit and switch pulses can mitigate this issue. This ensures that the coupling between the qubits is established only after the trajectory has passed the CZ operation point.} \label{fig:fig_gate_schematic}
\end{figure}

\begin{figure}[tb]
    \centering
    \includegraphics{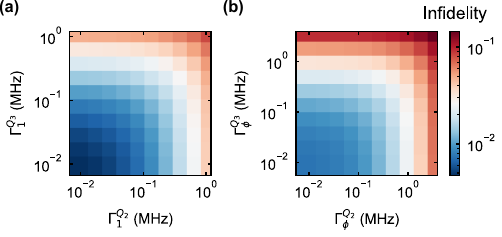}
    \caption{Numerical simulations of the CZ gate, using the parameters for $Q_{2}$ and $Q_{3}$ in Table~\ref{tab:qubit_params}. For the switches, we assume a reasonable decoherence time of $T_{1} = 10~\mu s$ and $T_{\phi} = 1~\mu s$, with an idle frequency of $1.67$ GHz. $\Gamma_{1}$ and $\Gamma_{\phi}$ indicate the decay rate and pure dephasing rate respectively. In (a), varying the qubit $T_1$ but not including qubit pure dephasing, a CZ gate fidelity as high as $99.5\%$ is achieved for qubit $T_{1}$ approaching $100~\mu s$. In (b), both qubits' $T_{1}$ are chosen from Table~\ref{tab:qubit_params}, with the pure dephasing rates varied along the horizontal axis. While the simulated fidelities with the measured $T_{\phi}$ are lower than the measured fidelity, we believe the discrepancy is due to the actual noise spectrum in the system. \label{fig:fig_cz_simu}}
\end{figure}

\begin{figure}[tb]
    \centering
    \includegraphics{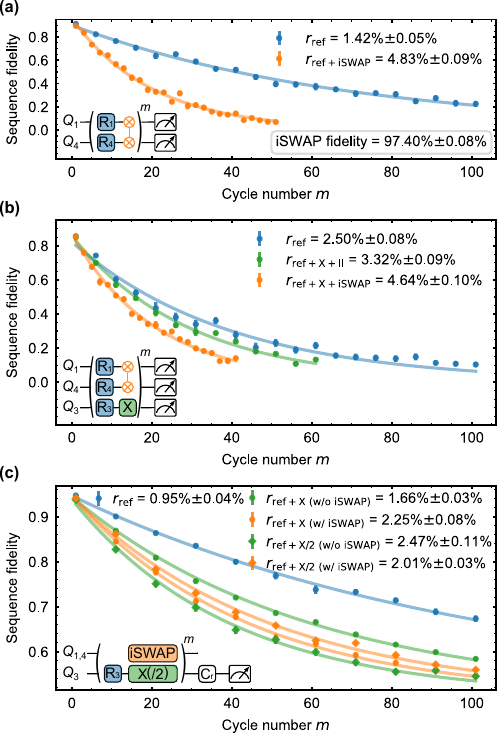}
    \caption{Benchmarking the iSWAP gate. (a) Cross-entropy benchmarking (XEB), measured by interleaving the iSWAP gate with random single-qubit gates $R_i$, yielding an iSWAP fidelity of $97.48\%\pm0.04\%$. (b) Performing XEB as in panel (a) while simultaneously performing interleaved single-qubit gates of a third qubit $Q_3$. (c) Randomized benchmarking (RB) of a $Q_3$ single-qubit gate with and without a $Q_1-Q_4$ iSWAP gate. Error bars and uncertainties of gate errors are generated by bootstrapping the experimental data, with $\pm1$ standard deviations shown in the plots.\label{fig:fig_iswap_exp}}
\end{figure}

\section{Tune-up procedure and evaluation of the CZ gates\label{sec:cz_tuneup}}
We implement the CZ gate with a resonant interaction between the two-qubit $\ket{ee}$ and $\ket{gf}$ (or $\ket{fg}$) states, where an $\pi$ phase is accumulated on $\ket{ee}$ within one round trip of the interaction (see Fig.~\ref{fig:fig_gate_schematic}(a)). We start with the two relevant qubits far detuned, then apply fast external flux bias pulses to the qubit SQUID loops such that the joint qubit $\ket{ee}$ state is brought close in frequency to either the $\ket{gf}$ or $\ket{fg}$ states, initiating a swap between the near-resonant states. The flux is then maintained for the appropriate duration such that the $\ket{ee}$ population performs a full swap and thereby acquires the desired $\pi$ phase shift. The flux bias is then adjusted to return the qubits to their idle frequencies. The process in terms of frequency trajectories is illustrated in Fig.~\ref{fig:fig_gate_schematic}(c), where as an example $Q_1$ is higher in idle frequency than $Q_2$, therefore they first approach the avoided crossing point between $\ket{ee}$ and $\ket{fg}$. Taking single-qubit phases due to frequency tuning into consideration, a perfect CZ gate has the matrix representation
\begin{equation}\label{eq:CZ_U}
\mathcal{U}_{\mathrm{CZ}}=
    \begin{pmatrix}
        1 & 0 & 0 & 0\\
        0 & e^{i\alpha_{2}} & 0 & 0\\
        0 & 0 & e^{i\alpha_{1}} & 0\\
        0 & 0 & 0 & -e^{i\left(\alpha_{1}+\alpha_{2}\right)}
    \end{pmatrix}
\end{equation}
on the basis states $\left\{ 
\ket{gg}, \ket{ge}, \ket{eg}, \ket{ee} \right\}$, where $\alpha_{1}~(\alpha_{2})$ denotes the single-qubit phase of the first (second) qubit. The calibration process of this CZ gate realization therefore involves three steps:
\begin{enumerate}
    \item Minimizing $\ket{ee}$ leakage. When we initialize the qubits to $\ket{ee}$, after the CZ gate, the population should remain in $\ket{ee}$ after interacting with $\ket{gf}$ or $\ket{fg}$. Loss in the $\ket{ee}$ population is classified as leakage into either the other energy levels of the two qubits, or into other modes such as the switch modes.
    \item Calibration of the controlled phase. This calibration is accomplished by first preparing the qubits in $\ket{g+}$, where $\ket{+} = \left( \ket{g} + \ket{e} \right) / \sqrt{2}$, evolving this state with the test CZ gate, then measuring the phase $\phi_{0}$ of the second qubit. The same procedure is repeated but instead with the qubits initialized in $\ket{e+}$, and the second qubit's phase $\phi_{1}$ measured. The conditional accumulated phase $\delta\phi=\phi_{1}-\phi_{0}$ then needs to be equal to $\pi$. As the CZ gate doesn't distinguish between control and the target qubits, this process also applies when initializing the qubits in $\ket{+g}$ and $\ket{+e}$, and measuring the conditional phase accumulated on the first qubit.
    \item Determining single-qubit phases. If we want to calibrate $\alpha_{2}$ in Eq.~(\ref{eq:CZ_U}), we simply prepare $\ket{g+}$ and let it evolve under the CZ gate. The phase of the second qubit then becomes $\alpha_{2}$. Similarly, $\alpha_{1}$ can be extracted by instead initializing in $\ket{+g}$ and measuring the terminal phase of the first qubit.
    \end{enumerate}

Following calibration, all gate parameters are stored for the subsequent XEB process. We repeat the same procedure for all six qubit pairs; Table~\ref{tab:CZ} shows a summary of the results.

It's also worth mentioning the procedure to extract the inferred CZ gate fidelity in Fig.~\ref{fig:fig3}(b). This is done by assuming the multiplicativity of the decay rate $p$ \cite{Arute2019} such that we have $p_{\mathrm{X}+\mathrm{CZ}}=p_{\mathrm{X}}\cdot p_{\mathrm{CZ}}$, where $p_{\mathrm{X}+\mathrm{CZ}}$ and $p_{\mathrm{X}}$ can be derived from panel (b) and (c) of Fig.~\ref{fig:fig3} respectively. Note that we also have the relation between the fidelity and the decay rate of any evolution $\mathcal{F} = 1 - (1-1/d)(1-p)$, therefore by plugging in the corresponding measured decay rates we are able to work out the inferred gate fidelity.

\begin{table*}[htbp]
 \caption{CZ gate table. An average XEB fidelity of $96.00\%\pm0.08\%$ is achieved with an average gate time of 46.1 ns. \label{tab:CZ}}
 \centering
 \begin{ruledtabular}
 \bgroup
 \def\arraystretch{1.5}
 \begin{tabular}{ccccccc}
   &$Q_{1}Q_{2}$&$Q_{1}Q_{3}$&$Q_{1}Q_{4}$&$Q_{2}Q_{3}$&$Q_{2}Q_{4}$&$Q_{3}Q_{4}$\\
   \hline
   XEB fidelity&$96.15\%\pm0.13\%$&$95.14\%\pm0.22\%$&$96.28\%\pm0.20\%$&$97.14\%\pm0.07\%$&$94.88\%\pm0.32\%$&$96.40\%\pm0.10\%$\\
   Gate time&37.0 ns&43.5 ns&43.5 ns&43.6 ns&61.5 ns&47.5 ns
 \end{tabular}
 \egroup
 \end{ruledtabular}
\end{table*}

To evaluate the performance of the CZ gate, we numerically simulate a two-node circuit with two qubits and two switches, with each mode modeled as a three-level system. The dynamics follow Eq.~(\ref{eq:me}), with all parameters evaluated using Eqs.~(\ref{eq:qubit freq}), (\ref{eq:qubit anh}) and (\ref{eq:coupling}) at each time step. All pulses are flattop with $5$ ns rising and falling edges, with a buffer time of $10$ ns between the qubit and switch flux pulses. The simulation results are shown in Fig.~\ref{fig:fig_cz_simu}, where we demonstrate that with reasonable switch coherence, the $T_{1}$ limit of the CZ gate can go beyond $99\%$, where in the experiment the pure dephasing of the qubits can likely be improved \cite{Hertzberg2021, Balaji2024, Hutchings2017}. Also, we find that lower-frequency switches tend to poison gate performance due to thermal occupancy of higher levels, implying that increasing the overall frequency of the system might improve the gate fidelity. Imperfections in the simulations may arise from use of RWA, as the low switch frequency will contribute to non-trivial fast rotating terms.

Other than directly simulating the system dynamics, Ref.~\cite{Abad2024} provides a formula for estimating the impact of decoherence on the fidelity of CZ gates, given by
\begin{equation}
    \overline{\mathcal{F}}_{\mathrm{CZ}} = 1 - \frac{1}{2}\frac{\tau}{T_{1}^{q1}} - \frac{3}{10}\frac{\tau}{T_{1}^{q2}} - \frac{31}{40}\frac{\tau}{T_{\phi}^{q1}} - \frac{3}{8}\frac{\tau}{T_{\phi}^{q2}},
\end{equation}
where $\tau$ is the gate time, and $q_{1}$ is the qubit which evolves to $\ket{f}$ during the gate. Using the parameters for $Q_{2}$ and $Q_{3}$ in Tables~\ref{tab:qubit_params} and \ref{tab:CZ}, we can work out the estimated CZ gate fidelity between them as $95.08\%$, consistent with our measurement results in Fig.~\ref{fig:fig3}.

\section{Tune-up procedure and evaluation of the iSWAP gates\label{sec:iswap_tuneup}}
We implement the iSWAP gate with a resonant two-qubit interaction between $\ket{eg}$ and $\ket{ge}$ (see Fig.~\ref{fig:fig_gate_schematic}(b)). An example of the dynamic trajectory of the energy levels is shown in Fig.~\ref{fig:fig_gate_schematic}(c). There are two factors to consider in addition to the single-qubit phases: First, the non-zero $ZZ$ interaction strength $\zeta$, discussed in Appendix~\ref{sec:impedance_response}. The effect of $\zeta$ is equivalent to a partial CZ gate, i.e. a controlled-phase gate with the phase determined by $\zeta$ and the gate duration. This adds a phase shift to the last diagonal entry of the evolution matrix. For the implementation of CZ gates, this effect can be easily compensated by detuning qubits away from their operation frequencies. For the iSWAP gates, this can be actively corrected by implementing partial CZ gates, or exploited to form other classes of gates. Second, the dynamical phase is due to the mismatch of the two qubits' idle frequencies (in other words, lack of synchronization of the two qubit rotating frames). This can be understood using a toy model involving only the two qubits. Similar to Eq.~(\ref{eq:H_rot}), the Hamiltonian for the two qubits in the rotating frame reads
\begin{equation}\label{eq:H_2q}
    \mathcal{H}_{2q} = \Delta_{1}a_{1}^{\dagger}a_{1} + \Delta_{2}a_{2}^{\dagger}a_{2} + g\left( e^{i\delta_{0}t}a_{1}^{\dagger}a_{2} + e^{-i\delta_{0}t}a_{1}a_{2}^{\dagger} \right),
\end{equation}
which assumes the rotating frames for the qubits coincide at $t=0$. Here $\Delta_{1}$ and $\Delta_{2}$ represent the detuning of each qubit, while $\delta_{0} = \omega_{0}^{1} - \omega_{0}^{2}$ denotes the difference in their idle frequencies. Using this Hamiltonian, we calculate the unitary evolution within a time interval as
\begin{equation}
    \mathcal{U}(t_{1}, t_{2}) = \mathcal{T}\exp(-i\int_{t_{1}}^{t_{2}}\mathcal{H}_{2q}\mathrm{d}t),
\end{equation}
where $\mathcal{T}$ is the time-ordering operator. Suppose a gate $\mathcal{U}$ with duration $\Delta t$ is implemented at $t_a = 0$ and $t_b$, so the dynamics of $\Delta_{1}$, $\Delta_{2}$, and $g$ are identical over the time intervals $(0,\Delta t)$ and $(t_b,t_b + \Delta t)$. Due to the rotating term in Eq.~(\ref{eq:H_2q}), we have
\begin{equation}
    \mathcal{U}(t_b, t_b+\Delta t) = e^{i\delta_{0} t_b a_{1}^{\dagger}a_{1}}\mathcal{U}(0, \Delta t) e^{-i\delta_{0}t_b a_{1}^{\dagger}a_{1}},
\end{equation}
which becomes trivial when either $\delta_{0} = 0$ \cite{Sung2021} or $\comm{\mathcal{U}(0, \Delta t)}{a_{1}^{\dagger}a_{1}} = 0$ (longitudinal gates like the CZ gate). When $\delta_{0} \ne 0$, even though we apply the same pulse shapes for all the parameters, the unitary evolution will depend on the initial time $t_b$. However, thanks to the flexibility of XEB, we can incorporate this effect in the ideal unitary matrix as the circuit depth increases so that the mismatch of two rotating frames is tracked precisely.

\begin{figure}[tb]
    \centering
    \includegraphics{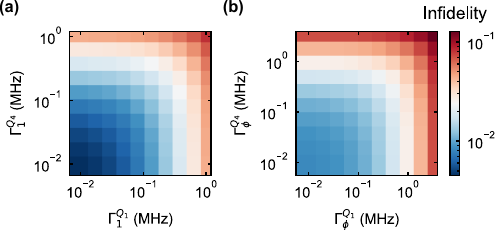}
    \caption{iSWAP gate numerical simulation, using the $Q_{1}$ and $Q_{4}$ parameters from Table~\ref{tab:qubit_params}. For the switches, we assume a reasonable decoherence time of $T_{1} = 10~\mu s$ and $T_{\phi} = 1~\mu s$, with an idle frequency of $1.67$ GHz. $\Gamma_{1}$ and $\Gamma_{\phi}$ indicate the decay rate and pure dephasing rate respectively. In (a), sweeping the qubit $T_1$ but with no qubit pure dephasing, an iSWAP gate fidelity as achieved $99.5\%$ is achieved with $T_{1}$ approaching $100~\mu s$. In (b), we vary the dephasing times while fixing both qubits' $T_{1}$ from Table~\ref{tab:qubit_params}. Differences between the simulated fidelities with the measured $T_{\phi}$ compared to the measured fidelities are likely due to the actual noise spectrum in the system. \label{fig:fig_iswap_simu}}
\end{figure}

\begin{figure}[tb]
    \centering
    \includegraphics{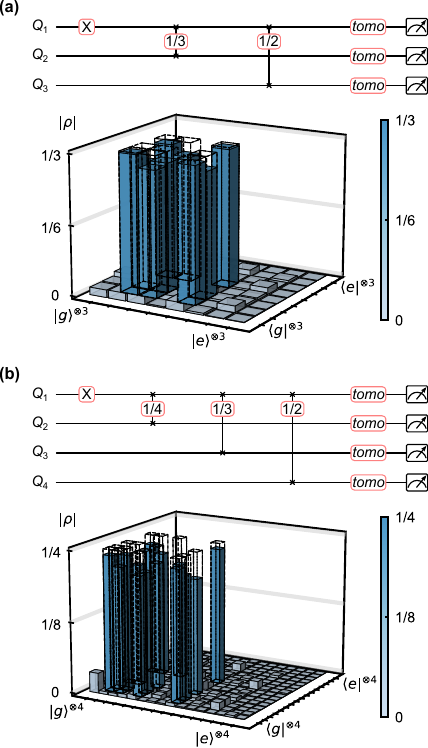}
    \caption{W state generation using a sequential partial swap procedure, with the corresponding circuit shown above each density matrix. In the circuit diagram, each value in the swap operation represents the portion of the excitation being transferred to the lower qubit. (a) Three-qubit W state with a fidelity $\mathcal{F} = 93.87\%\pm0.07\%$. (b) Four-qubit W state with a fidelity $\mathcal{F} = 88.87\%\pm0.08\%$. Fidelities are defined as $\mathcal{F} = \mathrm{tr}\left( \rho_{\text{ideal}}\rho \right)$. Error bars are $\pm 1$ standard deviation, extracted by bootstrapping the experimental data. \label{fig:fig_W}}
\end{figure}

\begin{figure}[tb]
    \centering
    \includegraphics{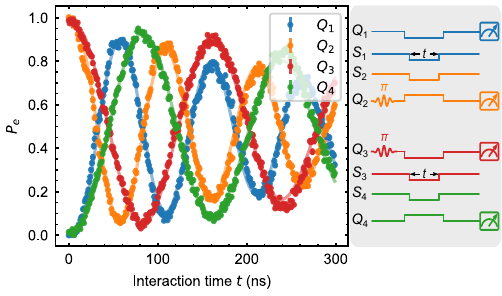}
    \caption{Simultaneous SWAP operations between $Q_{1}-Q_{2}$ and $Q_{3}-Q_{4}$, using swap frequencies separated by $220$ MHz. The pulse sequences are shown to the right, where we first excite $Q_{2}$ and $Q_{3}$, then bring $Q_{2}$ ($Q_{3}$) on resonance with $Q_{1}$ ($Q_{4}$), then turn on all switches for an interaction time $t$, followed by final multiplexed qubit measurements. \label{fig:fig_smtns_swap}}
\end{figure}

Taking both factors into account, we have a generalized matrix representation for an ideal $ZZ$-absorbed iSWAP gate \cite{Foxen2020}:
\begin{equation}
    \mathcal{U}_{\mathrm{iSWAP}} = 
    \begin{pmatrix}
        1 & 0 & 0 & 0\\
        0 & 0 & -ie^{i(\alpha_{2}-\delta_{0} t_{b})} & 0\\
        0 & -ie^{i(\alpha_{1}+\delta_{0} t_{b})} & 0 & 0\\
        0 & 0 & 0 & e^{i(\theta_{zz}+\alpha_{1}+\alpha_{2})}
  \end{pmatrix},
\end{equation}
where $\theta_{zz}$ denotes the conditional phase resulting from the $ZZ$ interaction.

There are three steps for calibrating the iSWAP gate:
\begin{enumerate}
    \item Tuning to resonance. We initialize the qubits in $\ket{eg}$, then perform the test iSWAP gate. We sweep one qubit frequency as well as the gate time to maximize the population in $\ket{ge}$ following the gate. As the iSWAP is symmetric for both qubits, these parameters should apply for the situation where we initialize in $\ket{ge}$ and end in $\ket{eg}$.
    \item Determining $\theta_{zz}$. The procedure is similar to measuring the $ZZ$ interaction strength in Fig.~\ref{fig:fig2}(d), using cross-Ramsey-type interference. We first prepare the qubits in $\ket{g+}$, then evolve under the iSWAP gate, following which the final phase of the first qubit is $p_{0} = \alpha_{1} + \delta_{0}t_{b} - \pi/2$. Next, we prepare the qubits in $\ket{e+}$, so that after the iSWAP gate the first qubit will accumulate a phase $p_{1} = \theta_{zz} + \alpha_{1} + \delta_{0}t_{b} + \pi/2$. Combining these we find $\theta_{zz} = p_{1} - p_{0} - \pi$.
    \item Determining the single-qubit phases. This is similar to the CZ gate process, except here there is the additional phase $\delta_{0}t_b$. To be more specific, to measure $\alpha_{1}$, we prepare the qubits in $\ket{g+}$, which will result in the first qubit picking up a phase of $\alpha_{1} + \delta_{0}t_b - \pi/2$; to measure $\alpha_{2}$, we prepare the qubits in $\ket{+g}$, which will result in the second qubit picking up a phase of $\alpha_{2} - \delta_{0}t_b - \pi/2$. In principle we know both $\delta_{0}$ and $t_b$, so both $\alpha_{1}$ and $\alpha_{2}$ can be obtained.
\end{enumerate}

\begin{table*}[htbp]
 \caption{iSWAP gate table. An average XEB fidelity of $96.60\%\pm0.05\%$ is achieved with an average gate time of 45.1 ns. We also list the XEB fidelities excluding the $ZZ$ phases, with an average fidelity of $94.36\%\pm0.14\%$.\label{tab:iSWAP}}
 \centering
 \begin{ruledtabular}
 \bgroup
 \def\arraystretch{1.5}
 \begin{tabular}{ccccccc}
   &$Q_{1}Q_{2}$&$Q_{1}Q_{3}$&$Q_{1}Q_{4}$&$Q_{2}Q_{3}$&$Q_{2}Q_{4}$&$Q_{3}Q_{4}$\\
   \hline
   XEB fidelity&$97.06\%\pm0.09\%$&$96.11\%\pm0.13\%$&$97.41\%\pm0.08\%$&$95.61\%\pm0.16\%$&$97.38\%\pm0.15\%$&$96.05\%\pm0.10\%$\\
   (no $ZZ$ phase)&$94.14\%\pm0.45\%$&$94.87\%\pm0.27\%$&$94.57\%\pm0.31\%$&$92.81\%\pm0.34\%$&$95.65\%\pm0.30\%$&$94.15\%\pm0.35\%$\\
   Gate time&49.9 ns&41.2 ns&43.4 ns&51.6 ns&47.0 ns&37.8 ns
 \end{tabular}
 \egroup
 \end{ruledtabular}
\end{table*}

Using the calibration and benchmarking methods mentioned above, we follow the same processes as in Fig.~\ref{fig:fig3} to benchmark the iSWAP gate between $Q_{1}$ and $Q_{4}$, with the results shown in Fig.~\ref{fig:fig_iswap_exp}. We benchmark the gate fidelity for all the qubit pairs, as summarized in Table~\ref{tab:iSWAP}, where we also include the results of the gate fidelity compared to a normal iSWAP gate excluding the $ZZ$ phase as a reference.

We numerically simulate the iSWAP gate in a two-node circuit (see Fig.~\ref{fig:fig_iswap_simu}) using the same framework as the CZ gate. We obtain a similar $T_{1}$ limit of the iSWAP gate, again assuming reasonable coherence of the switches. These simulations support the assumption that the gate fidelities are dominated by incoherent errors, which can likely be improved.

Similar to the CZ gate, Ref.~\cite{Abad2022} provides a formula for estimating the impact of decoherence on the fidelity of iSWAP gates, given by
\begin{equation}
    \overline{\mathcal{F}}_{\mathrm{iSWAP}} = 1 - \frac{2}{5}\tau\left( \frac{1}{T_{1}^{q1}} + \frac{1}{T_{1}^{q2}} + \frac{1}{T_{\phi}^{q1}} + \frac{1}{T_{\phi}^{q2}} \right).
\end{equation}
Plugging in the parameters of $Q_{1}$ and $Q_{4}$ in Tables~\ref{tab:qubit_params} and \ref{tab:iSWAP}, we can work out the estimated iSWAP gate fidelity between them as $95.70\%$, consistent with our measurement results in Fig.~\ref{fig:fig_iswap_exp}.

\section{W state\label{sec:W}}
In the main text, we show entanglement generation in the form of GHZ states. Here we display similar results for W states, another form of entanglement that is not interchangeable with GHZ states using local operations and classical communication (LOCC). The general form of an $N$-qubit W state is written as
\begin{equation}\label{eq:W}
    \ket{W_{N}} = \frac{1}{\sqrt{N}}\left( 
\ket{eg \cdots g} + \ket{ge \cdots g} + \ket{gg \cdots e} \right),
\end{equation}
which is symmetric upon the permutation of the excitation. Interesting properties of W states include robustness against excitation loss compared to GHZ states, as well as potential applications in approximate error correcting codes \cite{Faist2020}.

To generate the W states, we adopt the sequential partial swap method \cite{Hffner2005}, where the process starts with one excitation in the first qubit, i.e. the initial state $\ket{eg \cdots g}$. The excitation is then partially swapped to the second qubit with a probability of $(N-1):1$, which results in the state $\sqrt{(N-1)/N}\ket{eg \cdots g} + 1/\sqrt{N}\ket{ge \cdots g}$. The remaining excitation of the first qubit is then swapped to the $(i+1)$-th qubit with a ratio of $(N-i):1$, iterating $i$ until all qubits are occupied, resulting in the state in Eq.~(\ref{eq:W}). We use this process to generate three- and four-qubit W states, as shown in Fig.~\ref{fig:fig_W}, with fidelities sufficient to demonstrate genuine entanglement. Here, similar to the GHZ states, the density matrices are reconstructed using convex optimization methods, with the error bars extracted by bootstrapping.

\section{Simultaneous SWAP\label{sec:smtns_swap}}
The efficiency of a quantum processor can be boosted by implementing parallel gates. We have demonstrated the feasibility of performing single- and two-qubit operations at the same time (see Fig.~\ref{fig:fig4}); here we demonstrate that by sufficiently detuning two-qubit operation frequencies for different qubit pairs, we can realize simultaneous qubit swaps, with results shown in Fig.~\ref{fig:fig_smtns_swap}. The fidelity is estimated to be lower-bounded by about $90\%$ ($93\%$) for the $Q_1-Q_2$ ($Q_3-Q_4$) swap, which is calculated by comparing the first highest population of $Q_1$ ($Q_4$) to unity. One interesting and important fact to note is that as more switches are turned on at the same time, the pulse amplitudes needed to achieve a target qubit coupling tend to decrease. This is due to intra-switch coupling that increases the effective switch frequencies as more switches are biased to higher frequencies, an intrinsic feature of coupled multi-mode resonators.

\clearpage
\bibliography{ref}

\end{document}